\documentclass[12pt,a5paper,longbibliography]{article}
\usepackage[utf8]{inputenc}
\usepackage[usenames,dvipsnames]{xcolor}
\usepackage{amsmath}
\usepackage{amsthm}
\usepackage{setspace}
\usepackage{amsfonts}
\usepackage{booktabs}
\usepackage{multirow}
\usepackage{subfig}
\usepackage{enumitem}
\usepackage{graphicx}
\usepackage{changepage}
\usepackage{pdfpages}
\usepackage[bottom]{footmisc}
\usepackage[normalem]{ulem}
\useunder{\uline}{\ul}{}
\usepackage{lscape}
\usepackage[tableposition=below, labelfont=bf]{caption}
\usepackage[normalem]{ulem}
\useunder{\uline}{\ul}{}
\usepackage{array}
\usepackage[english]{babel}
\usepackage{natbib}
\usepackage{amssymb}
\usepackage[letterpaper, margin=1.25in]{geometry}
 
\usepackage{tikz}
\usetikzlibrary{calc}
\usepackage{booktabs}
\usepackage{caption}
\usetikzlibrary{decorations.pathreplacing}
\usepackage{pdfpages}
\usepackage{epigraph}
\usepackage{lscape}
\usepackage{multicol}
\setcitestyle{authoryear, round, citesep={;}}
\definecolor{ltgrey}{RGB}{180, 180, 180}

\usepackage{textcase}
\usepackage[T1]{fontenc}
\usepackage{fouriernc}
\makeatletter
\renewcommand\section{\@startsection{section}{1}{0mm}{-\baselineskip}{0.5\baselineskip}{\normalfont\centering\scshape}}
\renewcommand\subsection{\@startsection{subsection}{2}{0mm}{-\baselineskip}{0.5\baselineskip}{\normalfont\centering\itshape}}
\renewcommand\subsubsection{\@startsection{subsubsection}{3}{0mm}{-\baselineskip}{0.5\baselineskip}{\normalfont\centering\itshape}}
\def\@seccntformat#1{\@ifundefined{#1@cntformat}%
   {\csname the#1\endcsname.\ \ }      
   {\csname #1@cntformat\endcsname}}    
\newcommand\subsection@cntformat{\thesection.\Alph{subsection}. }
\newcommand\subsubsection@cntformat{\thesection.\Alph{subsection}.\arabic{subsubsection}. }   
\makeatother

\theoremstyle{plain}
    \newtheorem{result}{Result}
\theoremstyle{definition}

\theoremstyle{remark}

\renewcommand{\thesection}{\Roman{section}}

\usepackage{hyperref}

\begin{document}

\title{Revealed Incomplete Preferences\thanks{We thank Marina Agranov, David Ahn, G\'{a}bor B\'{e}k\'{e}s, Federico Echenique, Simone Galperti, Guillaume Fr\'{e}chette, Yoram Halevy, Paul J. Healy, Muriel Niederle, Lee Pang, Pietro Ortoleva, Charles D. Sprenger, Colin Sullivan, Emanuel Vespa, Ryan Webb, and Lanny Zrill for helpful comments and suggestions. This study was reviewed and granted exemption by the Institutional Review Board at California Institute of Technology. }}

\author{Kirby Nielsen\thanks{\texttt{kirby@caltech.edu}; Division of the Humanities and Social Sciences, California Institute of Technology} \hspace{0.1pt} \space \space \space Luca Rigotti\thanks{\texttt{luca@pitt.edu}, Department of Economics, University of Pittsburgh }}
\date{\today }
\maketitle

\singlespacing
\begin{abstract}

We elicit incomplete preferences over monetary gambles with subjective uncertainty. Subjects rank gambles, and these rankings are used to estimate preferences; payments are based on estimated preferences. About 40\% of subjects express incompleteness, but they do so infrequently. Incompleteness is similar for individuals with precise and imprecise beliefs, and in an environment with objective uncertainty, suggesting that it results from imprecise tastes more than imprecise beliefs. When we force subjects to choose, we observe more inconsistencies and preference reversals. Evidence suggests there is incompleteness that is indirectly revealed---in up to 98\% of subjects---in addition to what we directly measure.

\textsc{keywords:} incomplete preferences; subjective uncertainty; imprecise beliefs; imprecise tastes

\end{abstract}

\newpage

\singlespacing

\epigraph{It is conceivable and may even in a way be more realistic to allow for cases where the individual is neither able to state which of two alternatives he prefers nor that they are equally desirable.}{von Neumann and Morgenstern, \textit{Theory of Games and Economic Behavior}}
\vspace{-18pt}\epigraph{Is it rational to force decisions in such cases?}{Aumann, \textit{Utility Theory Without the Completeness Axiom}}
\onehalfspacing

Completeness is one of the most fundamental axioms on preferences; put simply, completeness states that an individual can rank any two alternatives. In contrast, when preferences are incomplete, there exist alternatives that the individual is unable to rank.\footnote{More formally, given a preference order $\succeq$, completeness states that between any two alternatives $p$ and $q$, either $p \succeq q$, or $q \succeq p$, or both; preferences are not complete if there exist $p$ and $q$ such that neither $p \succeq q$ nor $q \succeq p$.} Completeness is so central to standard economic theory that ``rationality'' is often defined as completeness plus transitivity (as, for example, in the very first definition of \citealp{mwg}), suggesting an implicit normative component to the axiom. Despite the widespread reliance on completeness in models of choice, there have been few attempts to measure the extent to which it is a valid assumption on preferences.\footnote{Notable early exceptions include \citealp{cohen-jaffray-said_85}, \citealp{cohen-jaffray-said_87}, and more recently  \citealp{cubitt2015imprecision}, \citealp{bayrak2020imprecision}, and \citealp{costagomes2021deferral}. We discuss these papers, among others, in Section~\ref{sec:literature}.} We introduce a new method for identifying incompleteness, and find that about forty percent of our subjects do not have complete preferences in a simple stochastic environment.

Incompleteness is inherently difficult to measure because decision-making experiments elicit \textit{choices} and typically force individuals to choose between alternatives. One could attempt to identify incompleteness by looking for indicators that might correlate with it, or for choice patterns that could be manifestations of individuals' attempts to complete their incomplete preferences. 
This often requires a model of how individuals complete their preferences, and whichever payment mechanism is imposed can muddy the interpretation of revealed incompleteness. For example, if one randomly picks an alternative when subjects report that they do not know which they prefer, then one could be capturing a preference for randomization rather than incompleteness of preferences.\footnote{This is especially an issue with subjective uncertainty since subjects can use randomization to eliminate ambiguity \citep{halevy2022randomize}.}

We try to circumvent some of these obstacles using a relatively simple new method for identifying incompleteness. We ask subjects to rank lotteries, and we use their rankings to ``estimate'' their preferences. Instead of paying subjects directly based on their rankings, we pay them based on what our estimated preferences predict they would choose in a question they never face. Subjects can skip questions when they are unable to rank the alternatives, and these questions will not enter into the preference estimation. Given these incentives, subjects can answer questions that would accurately inform the preference estimation, but need not answer questions when they ``do not know'' their preferred alternative. By indicating that they do not know how to rank the two lotteries, they directly reveal incompleteness. We discuss concerns of incentive compatibility with this procedure in Section~\ref{sec:design}, but our data suggest that subjects report truthfully and use the response options as we interpret them.

Our experiment reflects an interpretation of incompleteness that we believe is important for applications---we let individuals communicate that they do not believe some of their choices are welfare-relevant. As analysts, we often have a goal of observing an individual's choices and then using these choices to infer underlying preferences so that we can predict future choices, analyze counterfactual environments, make welfare comparisons, etc. Our methodology directly asks individuals to reveal the questions that they do not want us to use when we make inference about their underlying preferences. Observing incompleteness in this type of decision implies that individuals know the choices in which they are unsure about their preference, and suggests that the inability to reveal incompleteness could lead the analyst to inferences and predictions on behalf of the decision maker that are welfare-reducing.

We deploy this methodology in a simple online experiment with over 1,300 total subjects who compare monetary lotteries over a random binary event. We find that 39\% of our main sample express incompleteness in at least one comparison. Thus, a non-trivial minority of individuals have incomplete preferences. However, in any given binary choice, incompleteness is quite rare---individuals express incompleteness in only about 3\% of comparisons. Even subjects who do reveal some incompleteness do so in only about 7\% of comparisons. So while many individuals have incomplete preferences, the extent of incompleteness is small in our environment. Nevertheless, this incompleteness is systematic and largely conforms to simple predictions from theoretical formulations of incomplete preferences. Furthermore, we show that individuals report incompleteness for lotteries that are relatively ``far apart'' from one another, and that response times are slowest when individuals report incompleteness. We also allow individuals to report explicit indifference, and we show how incompleteness and indifference are distinct.

An inherent feature of our methodology is that we do not \textit{force} the subjects to make a choice. This is in contrasts with the standard ``Forced Choice'' paradigm of most decision-theoretic inspired experimental work where one asks subjects to choose between two lotteries without allowing them to express indifference or incompleteness. We compare the standard Forced Choice environment to our ``Non-Forced'' Choice environment using a within-subject elicitation that gives us the ability to detect the extent to which incompleteness could affect the inferences one makes about behavior in forced choice environments \citep{costagomes2021deferral}. In the Forced Choice treatment, each subject faces the same comparisons as in the Non-Forced treatment, but is asked to make choices without the option of expressing incompleteness or indifference. If preferences are incomplete but we do not give subjects the opportunity to reveal their incompleteness, then we should not be surprised when forced choices exhibit preference reversals or violations of basic properties like transitivity. We find that transitivity violations are far more common in the Forced treatment, and that incompleteness and indifference can explain about a third of these violations. 

Preference reversals could indicate that some individuals have underlying incompleteness that they are unaware of or that they do not reveal, and indeed some papers interpret preference reversals and randomization as incompleteness \citep{eliaz2006indifference, bayrak2017reversals}. We identify this by looking for cases of ``clear" preference reversals---comparisons in which subjects report a strict preference in the Non-Forced treatment and report the opposite preference in the Forced treatment. We refer to this as incompleteness that is ``indirectly revealed.'' 95\% of our subjects indirectly reveal incompleteness, and only 2\% of subjects neither directly nor indirectly reveal incompleteness. 

Interestingly, we find that the rates of directly and indirectly revealed incompleteness in any given question are highly correlated across the population. That is, the questions where subjects are most likely to exhibit preference reversals are the same questions in which other subjects are most likely to report incompleteness directly. This lends credibility to the interpretation of preference reversals as reflecting underlying incompleteness, and lends further credibility that our methodology accurately elicits the comparisons for which subjects are not sure of their preference. Furthermore, it suggests that some questions are inherently more ``difficult'' than others, which would be interesting to understand better in future work.

We design our experiment to reflect the growing theoretical literature on incomplete preferences in stochastic environments.\footnote{See, for example, \citealp{aumann1962completeness, bewley2002knightian, dubra2004euincomplete, eliaz2006indifference, gilboa2010objective, ok2012incomplete, galaabaatar2013seu, karni2022randomincomplete}}. This literature typically distinguishes between two possible sources of incompleteness: imprecise beliefs and imprecise tastes. To fix ideas, take the example of an individual who does not know which insurance policy they prefer. They might be unable to compare two policies because they are unsure how likely they are to fall ill in the coming year (imprecise beliefs), or because they are unsure of their risk tolerance (imprecise tastes). Our experiment mirrors this theoretical distinction by asking subjects to make choices between lotteries in a domain of subjective uncertainty. We design binary gambles specifying payoffs that subjects would receive if the Merriam-Webster Dictionary Word of the Day on a future date would be a verb or not a verb. Subjects can form a subjective belief about the likelihood that the word of the day will be a verb, but they might not be certain about this probability. This allows for incompleteness due to imprecise beliefs, and we identify belief imprecision by eliciting individuals' subjective belief as well as their certainty about this belief in the form of an unincentivized range of beliefs, following \citet{giustinelli2019precise}. 

47\% of subjects report uncertainty about their beliefs. Of these subjects, 43\% directly reveal incompleteness. However, among the remaining 53\% of subjects with certain beliefs, 36\% also directly reveal incompleteness. Thus, uncertainty in beliefs does not seem to be the primary source of incompleteness. Along the same lines, we find that individuals with larger ranges of imprecise beliefs are not more likely to report incompleteness. This suggests that imprecise \textit{tastes}---in our case, imprecise risk preferences----are the main source of incompleteness, rather than imprecise beliefs. We confirm this by running our exact same experiment with lotteries defined over an objective event. We tell subjects that the probability of the event (analogous to the likelihood that the word of the day would be verb) is equal to one third in one treatment and equal to one half in the other treatment. These objective probabilities mirror the modal beliefs from our subjective version of the experiment. We find that the same percentage of subjects report incompleteness in this treatment, and there is only a very small reduction of incompleteness at a comparison-level. Thus, we conclude that the main driver of incompleteness seems to be imprecise tastes rather than imprecise beliefs.

We see a few important implications of our results. First, incompleteness is important to understand for the reliability of two main goals of economics: assessing welfare and predicting behavior. Even though individuals might be \textit{able} to make choices when their underlying preferences are incomplete, this does not mean those choices should be used in welfare assessments. It is important to understand when individuals are unsure of their choice and would prefer it not to be used as indication of their preferences, which is exactly the motivation behind our elicitation mechanism. Related to this, a goal of estimating preferences is to predict future behavior. If individuals are unsure of their choice, it is likely the case that this choice and its implications on preference are not accurate predictors of future decisions. Identifying incompleteness in turn can yield better predictions.

Second, understanding the source of incompleteness can assist in targeting interventions to help individuals make decisions. Thinking back to our example of an individual who is unable to decide between two insurance plans, this could be because they are unsure about their beliefs, or it could be because they are unsure about their risk aversion. These two sources of incompleteness imply different policy interventions. In particular, belief uncertainty might call for targeted information provision. However, if incompleteness stems from preference uncertainty instead---as we find in our data---then these information interventions would go to waste. 

Furthermore, our objective lotteries are some of the simplest decisions we can ask individuals to make. We see incompleteness even in this environment, and it is natural to conjecture that incompleteness would only increase in more complicated settings. Indeed, we find that 76\% of subjects directly reveal incompleteness in a treatment where we make the choice objects more complex and ambiguous (a large increase from the 39\% of subjects in our main data). This suggests that the rates of incompleteness that we see in our main data represent a lower bound on the extent of incompleteness in choice. In addition, we find that forced choice leads to less-coherent preferences even in this simple environment, and we conjecture that forced choice would lead to even more inconsistencies in more complex environments. We leave further study of this to future work, but it would be interesting to extend analysis into more naturalistic choice environments with more complex comparisons. 

That said, we view our results as a positive message for researchers conducting simple experiments to measure and understand preferences. Incompleteness does exist and we can make cleaner inference on preferences when we allow subjects to express their incompleteness. However, incompleteness is relatively uncommon in these decisions, so forced choice paradigms are likely to approximate underlying tastes relatively well. 

Finally, as we discuss more in Section~\ref{sec:discussion}, our data and design open interesting questions on the nature and interpretation of incompleteness. We still observe preference reversals in our Non-Forced data. Some of this could be attributed to errors or stochastic preferences, but we find evidence that individuals have some incompleteness in their preferences that they are unaware of. Along similar logic, it could be that individuals approach our environment by developing heuristics or procedures in order to complete their preferences. We cannot identify this directly, but if this were the case, it provides another avenue by which our experiment underestimates the amount of incompleteness in preference. It would be very interesting to better understand the process of how individuals complete their preferences when forced to make choices.

\section{Theories of Incompleteness} \label{sec:theory}

A model of decision making under subjective uncertainty is presented in \cite{bewley2002knightian}. It shows that a strict preference relation that is not necessarily complete, but satisfies all other axioms of the standard Anscombe-Aumann framework, can be represented by a unique utility index and a \textit{set} of probability distributions. In this model, lack of completeness is thus reflected in multiplicity of beliefs: the unique subjective probability distribution of the standard expected utility framework is replaced by a set of probability distributions.

Since we restrict attention to monetary outcomes that depend on a binary event, we can denote the state as $\{v,nv\}$ (corresponding to ``verb'' and ``not verb'') and describe the set of all probability distributions over this state space using the interval $[0,1]$ with generic element $\pi$. Let $\succsim$ denote an individual's preference relation over elements of $\boldsymbol{R^2}$ (these are pairs of monetary outcomes in each state). Bewley's Knightian Decision Theory can be summarized by saying that for any $p,q \in \boldsymbol{R^2}$
\begin{equation}\label{equ:Knightian}
p\succsim q \qquad \text{if and only if}\qquad E_{\pi }[u(p)] \geq E_{\pi }[u(q)] \text{ for all }\pi \in \Pi
\end{equation}
where we define $E_{\pi }[u(p)] \equiv \pi u(p_v) + (1-\pi) u(p_{nv})$, and let $u(x)$ denote the utility yielded by the monetary amount $x$, and let $\Pi \subseteq[0,1]$. That is, $p$ is preferred to $q$ if it has higher expected utility for all probability distributions corresponding to the interval $\Pi$. If the inequality in (\ref{equ:Knightian}) changes direction for different elements of $\Pi$, the two alternatives are not comparable. In this model, alternatives are evaluated one probability distribution at a time, and they can be ranked only when all those evaluations agree. In this theory, the incompleteness is reflected by multiple subjective probabilities, and if $\Pi$ reduces to a singleton the preferences are complete and this is standard subjective expected utility.\footnote{See \cite{rigottishannon2005} for a precise statement of this result.}


 \cite{aumann1962completeness} models incompleteness through imprecise tastes. He proposes a model of incompleteness with objective lotteries that weakens the original von-Neumann \& Morgenstern axioms by dropping completeness.\footnote{Aumann's approach has been extended and clarified in \citealp{ok2002uincomplete, dubra2004euincomplete, eliaz2006indifference, ok2012incomplete}.} In that case, preferences are described by a single objective probability distribution and many utility functions. We can summarize Aumann's theory by saying that for any $p,q \in \boldsymbol{R^2}$
\begin{equation}\label{equ:Aumann}
p\succsim q \qquad \text{if and only if}\qquad E_{\pi }[u(p)] \geq E_{\pi }[u(q)] \text{ for all } u \in \mathcal{U}
\end{equation}
where $\mathcal{U}$ is a set of utility functions for monetary amounts. That is, $p$ is preferred to $q$ if it has higher expected utility for all utility functions. If the inequality in (\ref{equ:Aumann}) changes direction for different elements of $\mathcal{U}$, the two alternatives are not comparable. In this model, alternatives are evaluated one utility function at a time, and they can be ranked only when all those evaluations agree. Whenever $\mathcal{U}$ contains a single utility function, this model reduces to von-Neumann \& Morgenstern expected utility with objective probabilities.

For future reference, we label incompleteness described by multiple probabilities as ``imprecise beliefs'' theory and incompleteness described by multiple utilities as ``imprecise taste'' theory. These two theories are composed in \cite{galaabaatar2013seu} where one can have both imprecise beliefs as well as imprecise taste. This model can be described by the following representation
\begin{equation}\label{equ:GalaKarni}
p\succsim q \qquad \text{if and only if}\qquad E_{\pi }[u(p)] \geq E_{\pi }[u(q)] \text{ for all } u \in \mathcal{U} \text{ and } \pi \in \Pi
\end{equation}
This model allows for both imprecise beliefs and imprecise taste at the same time, and admits the previous two as special cases. 

We design our experiment to investigate these two sources of incompleteness. Our main environment is one of subjective uncertainty, allowing for incompleteness to result from imprecise beliefs and/or imprecise tastes. We attempt to disentangle these two channels in two ways. First, we classify subjects as having precise or imprecise beliefs based on a self-reported measure of belief precision. We say that, for subjects with precise beliefs, incompleteness can only result from imprecise tastes. Second, we exogenously eliminate belief imprecision in a treatment with objective uncertainty. This allows us to rule out incompleteness due to imprecise beliefs, leaving only incompleteness that results from imprecise tastes.

\section{Experimental Design} \label{sec:design}
We designed our experiment with three broad goals: 1. capture the extent of incompleteness and indifference in individuals' preferences, 2. analyze how forced choice affects inference on preferences relative to non-forced choice, and 3. identify the extent to which incompleteness results from imprecise beliefs relative to imprecise tastes. We conduct our experiment in the domain of simple binary choices over monetary lotteries. We chose this domain precisely for its simplicity and because questions of this form have been studied extensively in the literature, so our results can provide an easy comparison to previous work.

We first describe the lottery choices and treatments and then describe our method of eliciting incompleteness.

\subsection{The Event}
We chose a novel event structure over which to design our lotteries. The payoff of each lottery was determined by the part of speech of the Merriam-Webster Dictionary Word of the Day on a pre-specified future date.\footnote{The Merriam-Webster Dictionary posts a ``word of the day'' every day, intended to teach people new words.} For example, one lottery would pay \$2 if the word of the day in 3 days is a verb, and would pay \$14 if the word of the day in 3 days is not a verb. We provided subjects with a list of previous words of the day and their parts of speech for a past month to give them a sense of the frequency of each part-of-speech. We do not provide subjects with the empirical frequency of these parts of speech for either the Merriam-Webster Word of the Day or for the English language in general, and this was not easily found on the internet to the best of our knowledge.\footnote{In the list that we provided to subjects, 10 out of 30 words were verbs. This relates to why we calibrated our lotteries to target a region of incompleteness around one-third. As shown in our results, a large mode of subjects do report subjective beliefs near one-third.} 

At the end of the experiment, we elicit subjects' subjective belief that the word of the day on the pre-specified future date would be a verb. We discuss payment from the experiment below, but if subjects were paid for their beliefs, then we incentivized this report using an incentive-compatible Becker-DeGroot-Marschak procedure \citep{becker1964bdm}. Following this, we ask subjects whether they are \textit{certain} of this belief or not, and they respond with a simple yes or no answer. If they answer ``no,'' then we allow them to specify a range of beliefs in addition to their point estimate. We followed the elicitation procedure from \citet{giustinelli2019precise}, who elicit precise and imprecise beliefs about individuals' likelihood of developing late-onset dementia. This elicitation was unincentivized.\footnote{\citet{karni2018mechanism} and \citet{karni2020comparative} provide interesting elicitation methods for eliciting sets of subjective probabilities. To the best of our knowledge, these have not been validated behaviorally yet, and it would be interesting to compare these mechanisms to unincentivized reports. Given that our experiment was already quite complex, we use the simpler procedure from \citet{giustinelli2019precise} and rely on exogenously inducing certain beliefs to confirm our results.}

There are a few reasons why we used a subjective event in our main experiment. First, one of our goals in the paper is to understand the role of imprecise beliefs versus imprecise tastes in generating incompleteness in preferences. Therefore, we chose an event that was likely to lead to some belief imprecision. Second, most experiments that try to measure incompleteness use objective uncertainty (we review these papers in Section~\ref{sec:literature}). Thus, a contribution of our paper is to study incompleteness over simple monetary gambles, but allowing for individuals to form their own---potentially uncertain---subjective beliefs. This also allows us to compare incompleteness in objective versus subjective environments using the same experimental setup as we discuss below. 

We chose this specific event for a few reasons. Especially because we ran the experiment online, we wanted an event that was not controlled by the experimenter. We felt this might help subjects trust the ambiguous process of determining the state, without worrying that the experiment was ``rigged'' in some way. We also felt this would avoid issues of ``comparative ignorance'' or other concerns about the experimenter knowing the state while subjects did not \citep{fox1995comparative}. Furthermore, we wanted an event where uncertainty would reasonably lead to belief imprecision rather than invoking something like the principle of insufficient reason. Many ambiguous events result in subjects forming uniform beliefs (e.g., whether the temperature in a city across the world will be above or below $X$\ degrees), and we wanted to allow for more heterogeneity in belief formation.

\subsection{Lotteries}
\begin{figure}[!htb]
    \centering
    \includegraphics[width=\textwidth]{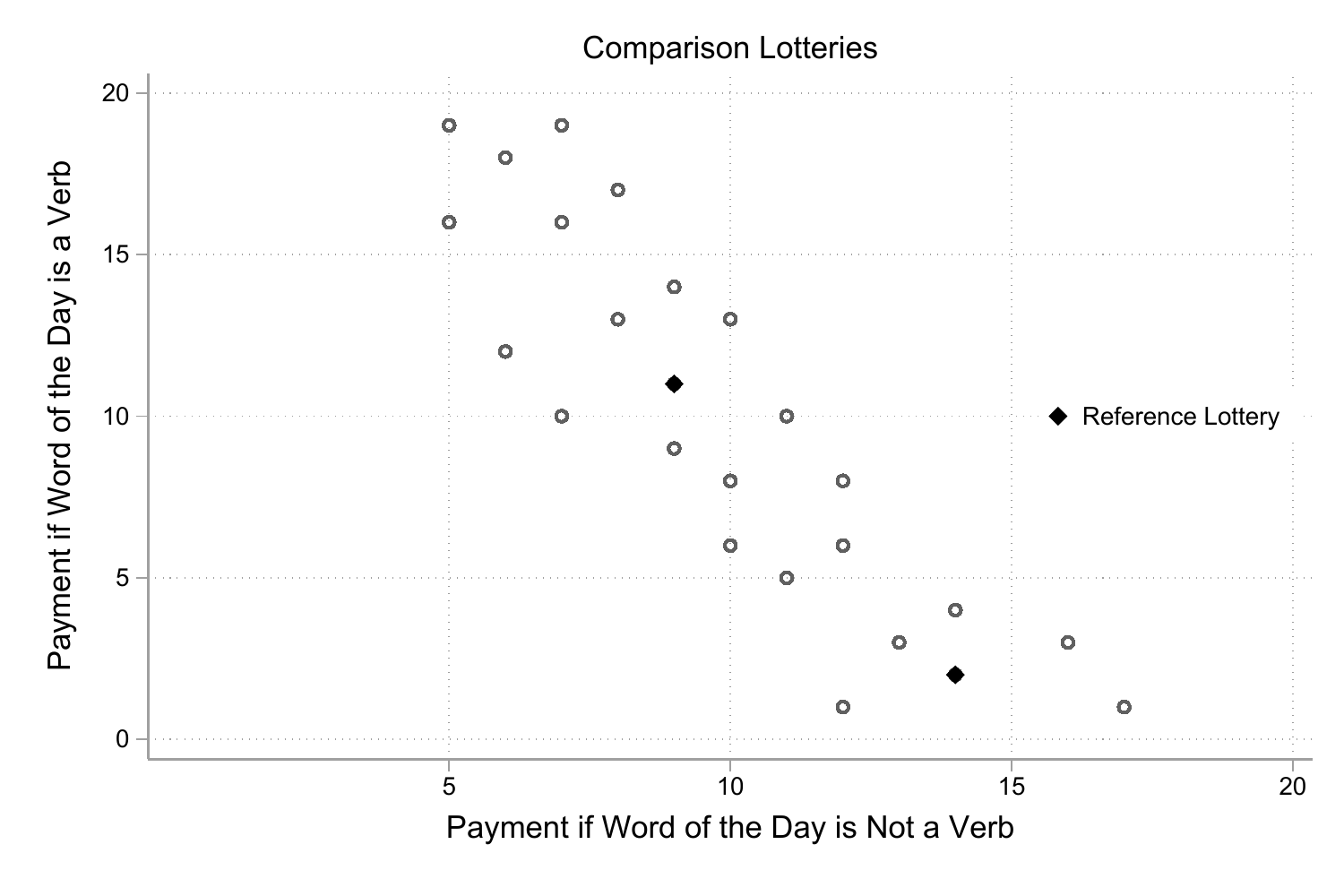}
    \caption{Comparison Lotteries}
    \label{fig:lotteries}
    \footnotesize{\underline{Note:} Points show the payoffs associated with the 25 lotteries used in the experiment. The black diamonds show the reference lotteries, $(9,11)$ and $(14, 2)$. Both reference lotteries were themselves comparison lotteries, so we had subjects compare each reference lottery to itself, to the other reference lottery, and to the 23 lotteries represented by the open circles.}
\end{figure}

Subjects make binary choices over lotteries that specify payoffs to be received if the word of the day is a verb or not a verb. We denote a lottery by $(nv, v)$ where \$$nv$ is the payoff the subject would receive if the word of the day is not a verb and \$$v$ is the payoff they would receive if the word of the day is a verb. There are two within-subject treatments (described below). In each treatment, subjects faced two blocks of 25 questions each. In a block, subjects compare 25 lotteries to a ``reference lottery'' for a total of 25 comparisons per block. \autoref{fig:lotteries} shows the two reference lotteries---(9, 11) and (14, 2)---and the 23 other lotteries used in the experiment; we also list all lotteries in Table~\ref{tab:listoflotteries} in the Appendix. We chose one reference lottery to be fairly symmetric and the other asymmetric across states. Note, the reference lotteries were themselves comparison lotteries, so we asked subjects to compare each reference lottery to itself, and asked subjects to compare the two reference lotteries to each other.

We chose these specific lotteries, rather than randomly-generated lotteries or other methods of selection, to target a region of likely incompleteness. In particular, we considered a range of beliefs around $pr(verb)=\frac{1}{3}$ with linear and log utility functions. For reference, this is visualized in the Appendix in Figures \ref{fig:142functions} and \ref{fig:911functions}. In addition to targeting incompleteness, we included some lotteries related by dominance, and some that would be comparable by most preferences.

To subjects in the experiment, we referred to the lotteries as ``gambles.'' We did not make a distinction between reference and comparison gambles. Instead, we said that one gamble would stay the same across a block while the other varied.

\subsection{Treatments}
We had two within-subject treatments. The ``Non-Forced'' treatment allowed subjects to report strict preference, indifference, and incompleteness. Specifically, subjects could report one of four options, reported verbatim below:
\begin{enumerate}[noitemsep]
    \item I rank Gamble 1 above Gamble 2
    \item I rank Gamble 2 above Gamble 1
    \item I rank Gambles 1 and 2 exactly the same
    \item I don't know how I rank Gambles 1 and 2 
\end{enumerate}
We interpret the first two options as strict preference, the third as indifference, and the fourth as incompleteness. The order of the four options on subjects' screens was randomized independently across each question. We include a screen shot in Appendix Figure~\ref{fig:screenshot}.

In the Forced treatment subjects could report one of two options:
\begin{enumerate}[noitemsep]
    \item I rank Gamble 1 above Gamble 2
    \item I rank Gamble 2 above Gamble 1 
\end{enumerate}
While we could have retained the indifference option in the Forced treatment, we intended this treatment to mirror the vast majority of experimental elicitations which do not include the option to report indifference. Again, the order of the options was randomized independently in each question. We told subjects that if they did not know how they ranked the gambles, or if they ranked them exactly the same, then they should ``choose one of the two possibilities that (they) think fits best.''

Subjects saw the Forced and Non-Forced treatments in random order. Within each treatment, we randomized the order of the two reference lottery blocks. Within each block, we randomized the order of the 25 lotteries. Thus, in total, subjects made 100 binary choices between gambles (25 lotteries $\times$ 2 reference lotteries $\times$ 2 treatments).

\subsection{Payment}
Subjects were paid for one random decision that they made throughout the experiment. There were three main payment possibilities: 1. a decision in the Non-Forced treatment, 2. a decision in the Forced treatment, or 3. their reported belief.

If they were paid for a decision in the Non-Forced treatment, then we implement our estimation procedure. We describe this in detail below. If they were paid for a decision in the Forced treatment, then we randomly selected one of the lottery choices and paid them the lottery they chose in this decision. This is an entirely standard incentivization for binary choices, so subjects should choose their preferred lottery in each binary decision. If they were paid for their reported belief, then we paid them according to the BDM procedure, for a bet worth \$5. 

Since payments were based on the word of the day in the future, subjects did not receive event-based payments immediately. They received their \$5 show-up fee on the day of the experiment, but, if randomly-selected to receive a bonus payment, then they received their bonus payment on the pre-specified future date 3 days in the future.\footnote{While this introduces a role for time preferences, there is no natural way for this to interact with the elicitation of incompleteness.} 

\subsection{The Algorithms}
To elicit incompleteness, we could simply ask individuals to tell us when they do not know their preference. If they answer carefully and honestly, this would give us a perfect measure of incompleteness. Without incentives to answer carefully and honestly, though, we might doubt that individuals' responses reflect their underlying preferences. We attempt to keep this simple and straightforward elicitation procedure---asking individuals to tell us when they do not know---but introduce some incentives to answer carefully and honestly.   

To do this, we adapt the basic idea of the elicitation used in \citet{krajbich2017neurometrically} and \citet{kessler2019resume}, adjusted to allow for the elicitation of incomplete preferences. In particular, we tell subjects that their responses will teach an ``algorithm'' the kinds of gambles that they prefer, and the algorithm will use these responses to pay them for a \textit{different} question at the end of the experiment. The question paid is not a question that was asked to subjects directly. The main idea is that reporting strict preference or indifference will help the algorithm to better understand a subject's preferences over gambles, but if a subject reports incompleteness, then this question does not enter into the algorithm. 

Below we overview the two different ``algorithms'' we implement, while their detailed description is in Appendix Section~\ref{sec:appendix_algorithms}. For both algorithms, following \citet{krajbich2017neurometrically}, \citet{kessler2019resume} and \citet{danz2022belief}, we give subjects minimal details about how the algorithm works. Specifically, we tell subjects the following: 
\begin{quote}
\begin{singlespacing}
We will not actually pay you directly for the gambles in these question groups. Instead, we will pay you based on what your responses imply about what gambles you prefer in some other decision that you will not face. We will use your choices in this question group to understand what types of gambles you like or dislike. At the end of the experiment, we will pick two gambles. We will pay you the gamble that we think you prefer out of those two randomly selected gambles. We will use your earlier choices to decide which gamble we think you prefer in this decision...
\begin{itemize}
    \item If you say that you rank one gamble over the other, then we will use this information to help our algorithm understand which gambles you would rather have.
    \item If you say that you rank the two gambles equally, then we will use this information to help our algorithm understand when having either of two gambles is the same to you.
    \item If you say that you do not know how to rank the two gambles, then we will not use that question in our algorithm.
\end{itemize}
\end{singlespacing}
\end{quote} 

We give subjects an example of how stating a preference for \$5 over \$4 can teach the algorithm that you prefer more money to less. Subjects can click a button to learn more detailed information about the algorithm; 27\% of subjects click this button, and these subjects are $\sim$7 percentage points more likely to report incompleteness.

One of our algorithms is a standard maximum likelihood estimation procedure. We fix a single payment question which is distinct from the questions that the subjects face in the experiment. We take all of the questions in which a subject reports strict preference or indifference and use these to estimate the risk aversion parameter in a CRRA utility function. The questions in which a subject reports incompleteness do not enter into this estimation. Then, we use this estimated utility function to predict which lottery the subject would prefer in our fixed payment question, and pay them based on this prediction. 

Under this algorithm, reporting a strict preference or indifference can help the maximum likelihood estimation procedure form a more precise estimate of the subject's risk parameter. When subjects are not sure of their preference in a given comparison, reporting incompleteness prevents the question from potentially biasing the estimation. 

However, this type of procedure forces complete preferences in the payment questions. Additionally, it relies on assuming a specific functional form and the mapping from choices to the payment question in light of this class of utility functions. Because of this, in other treatments, we also use a very different type of algorithm, based on a non-parametric construction of better-than and worse-than sets. These sets start with randomly-generated lotteries which are replaced via dominance whenever subjects report a strict preference or indifference. For example, if a subject reports preferring a lottery $p$ over the reference lottery, then we replace one of the lotteries in the better-than set by $p'$ which dominates $p$. We pay subjects based on a randomly-selected lottery from the better-than or worse-than sets. 

This procedure does not force complete preferences in an ex-ante specified payment question, since the question we ultimately pay subjects is one where we are ``sure,'' under dominance, that they have complete preferences. It also does not rely on any parametric assumptions on preferences. However, one major drawback of this type of procedure is that the questions subjects answer influence the possible lotteries they could be paid, since the lotteries that are replaced into the better-than and worse-than sets are a function (via dominance) of the questions in which a subject reports strict preference or indifference. 

In the end, both these estimation procedures have pros and cons. We ran both as a measure of robustness, and we find that subjects' responses do not depend on the details of the elicitation algorithm; see Appendix Section~\ref{sec:appendix_algorithms} for details. Given this insensitivity to the algorithms, despite their different incentives, we do not believe the details of the estimation affect how subjects respond. We do not advocate for a specific preference estimation algorithm to be used to elicit incompleteness, but rather we highlight the general idea of using choices to estimate preferences, while incentivizing the elicitation of incompleteness as the choices subjects do not believe to be representative of their preferences. 

We cannot rule out that subjects do not understand these algorithms or form beliefs about them such that manipulation is profitable. Because of this, we designed our experiment to be able to detect manipulation in a few ways. We conduct the Forced choice treatment within-subject, and this treatment is incentivized in the standard way, so we can compare these responses to those in the Non-Forced treatment. We also include specific questions to detect manipulation and specific treatments to test whether choices under the algorithm respond in the predicted ways. The overwhelming summary of our evidence suggests that subjects are reporting truthfully. 

Furthermore, we asked subjects why they (n)ever reported incompleteness. For those who never reported incompleteness, 91\% said this was because they ``always preferred one over the other.'' For those that did report incompleteness, 75\% said that they did so because they ``didn't know which gamble (they) preferred'' and 17\% said they ``did not know how to compare the gambles.'' We discuss these survey responses more in Section~\ref{sec:results}, and we organize all of the results that we use to validate the elicitation in Section~\ref{sec:manipulation}.

Taken together, these results suggest that our elicitation falls under the umbrella of mechanisms that are \textit{behaviorally} incentive compatible \citep{danz2022belief} even if not theoretically incentive compatible for all possible beliefs and preferences. We liken this to other methodologies such as dynamically optimized sequential experimentation (DOSE, \citealp{Wang2010DOSE}) and the other papers in the literature that have used preference estimation without entirely fixing subjects' beliefs about the incentives (\citep{krajbich2017neurometrically, kessler2019resume}).

\subsection{Participants}
In our main analysis, we analyze data from a total of 639 participants recruited through Prolific, an online participant recruitment platform.\footnote{We restricted to participants of US nationality, with at least 98\% approval rate, who had participated in at least 50, but no more than 500, previous studies on Prolific. These 639 subjects were collected in three waves. In May 2021, we recruited 119 subjects. We self-replicated our exact experiment, recruiting 382 additional subjects in November 2021. Results from the two waves are statistically indistinguishable, so we pool the data. We recruited 138 additional subjects in August 2022 using a different elicitation algorithm, which we discussion Appendix Section~\ref{sec:appendix_algorithms}. As shown in the Appendix, these results are also statistically indistinguishable, so we combine all three waves for our main analysis.} We paid subjects \$5 for completing the experiment. Subjects took 35 minutes to complete the experiment on average, so the $\sim$\$9/hr wage is commensurate with other studies on Prolific where researchers are required to pay at least \$6.50/hr. In addition, each subject had a 10\% chance of being randomly selected to receive a bonus payment based on their decisions in the experiment. The average bonus payment among those who received a bonus was \$12.10.

\section{Results} \label{sec:results}
\begin{figure}[!htb]
    \centering
    \centerline{\includegraphics[width=0.6\textwidth]{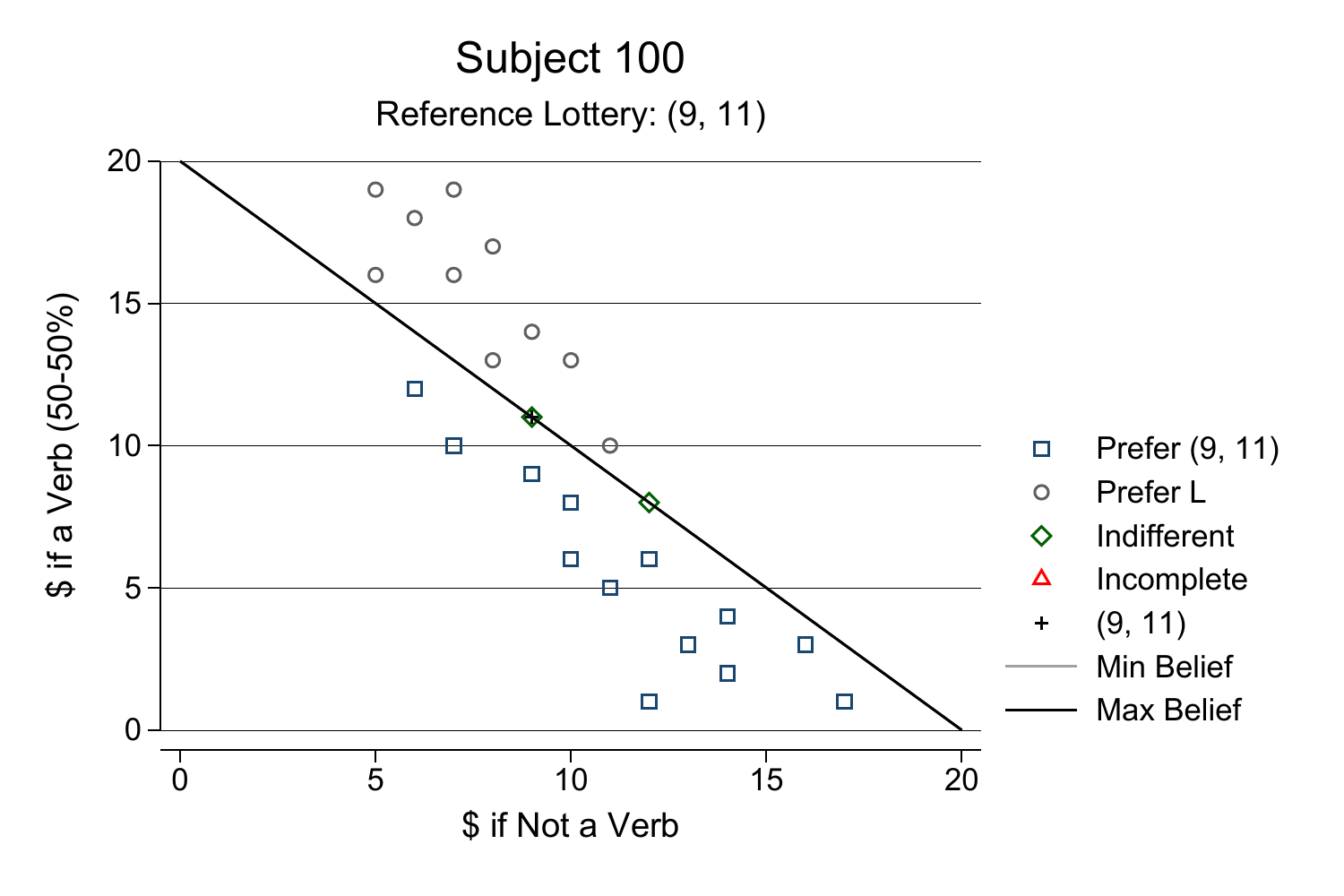}
    \includegraphics[width=0.6\textwidth]{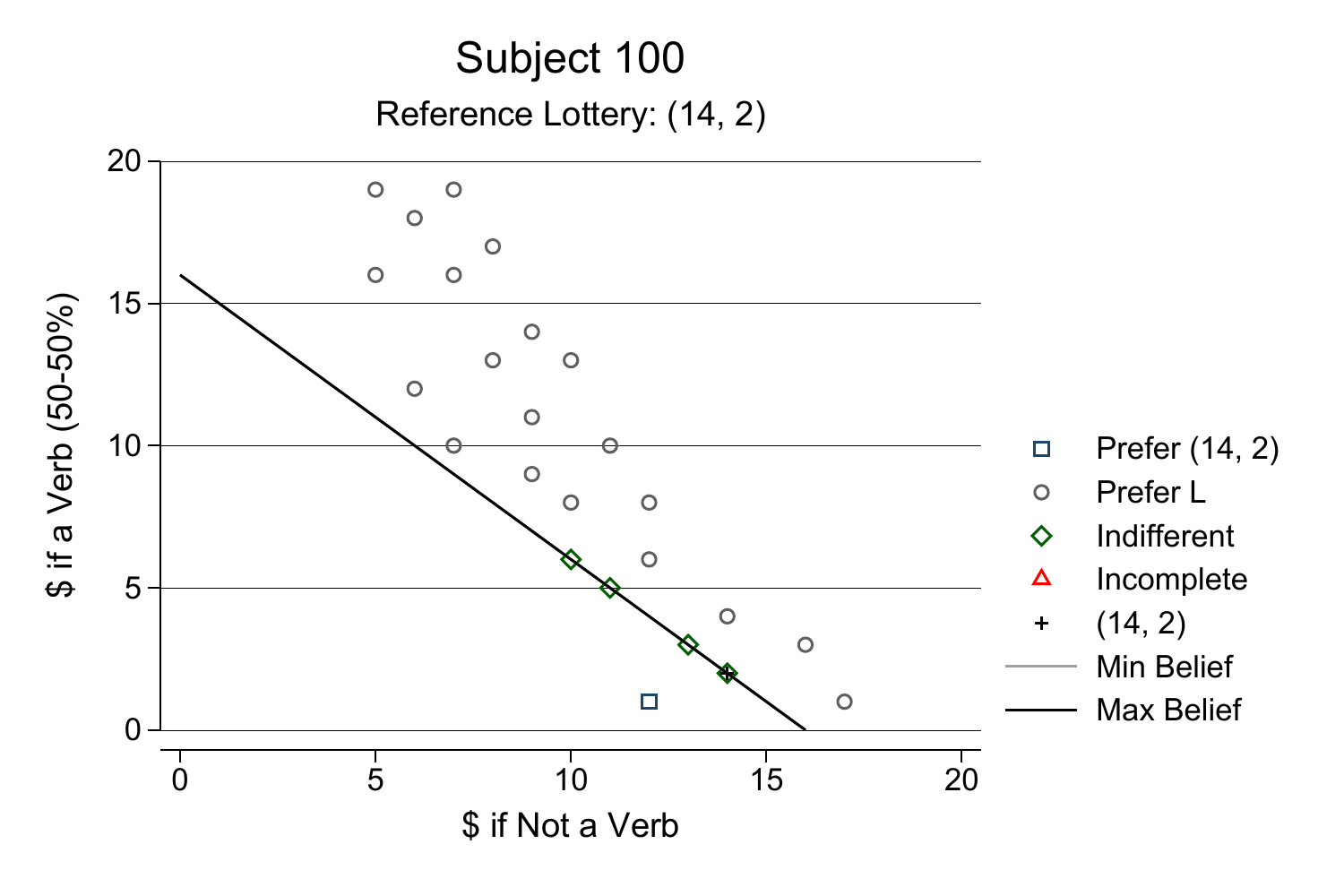}}
    \centerline{\includegraphics[width=0.6\textwidth]{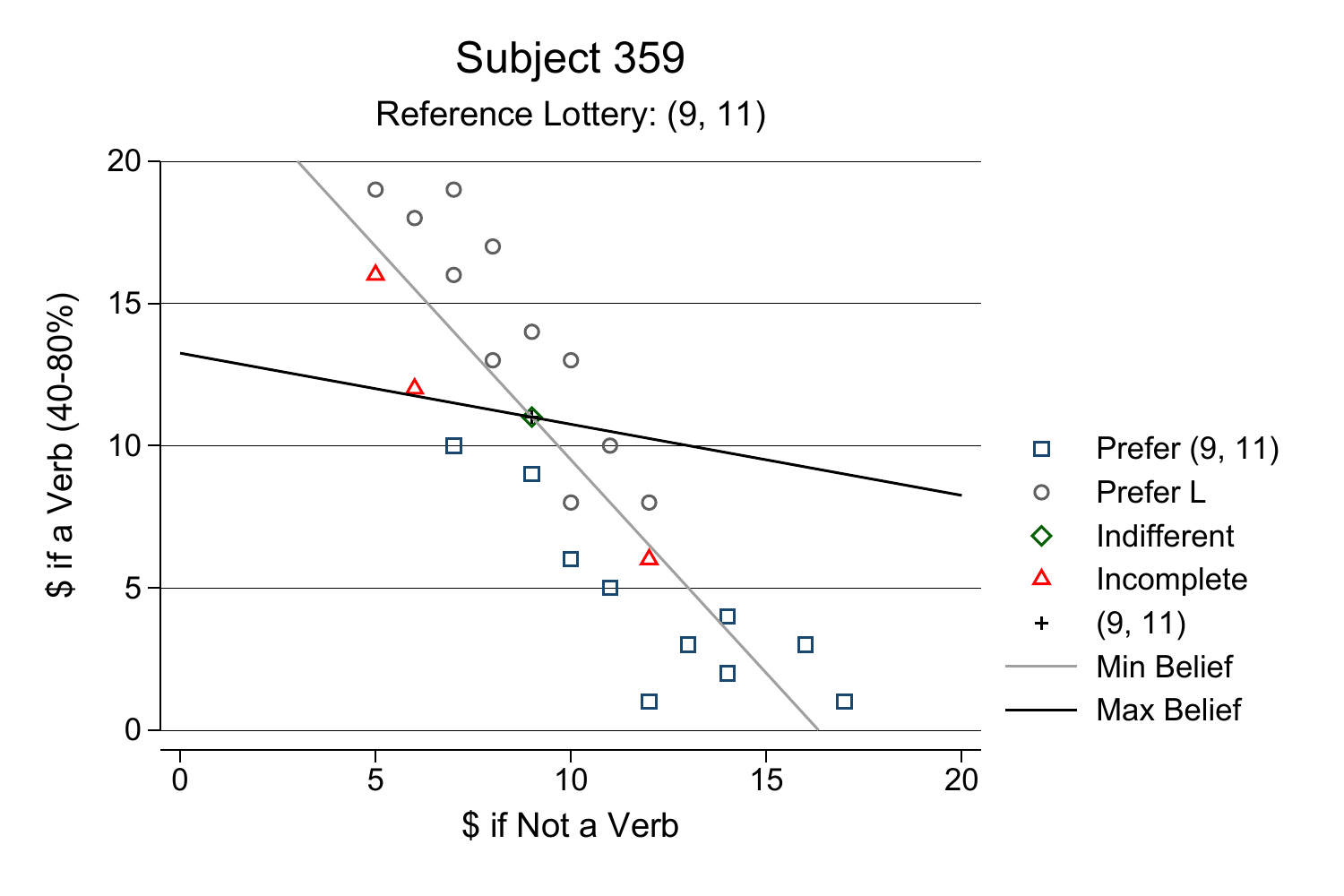}
    \includegraphics[width=0.6\textwidth]{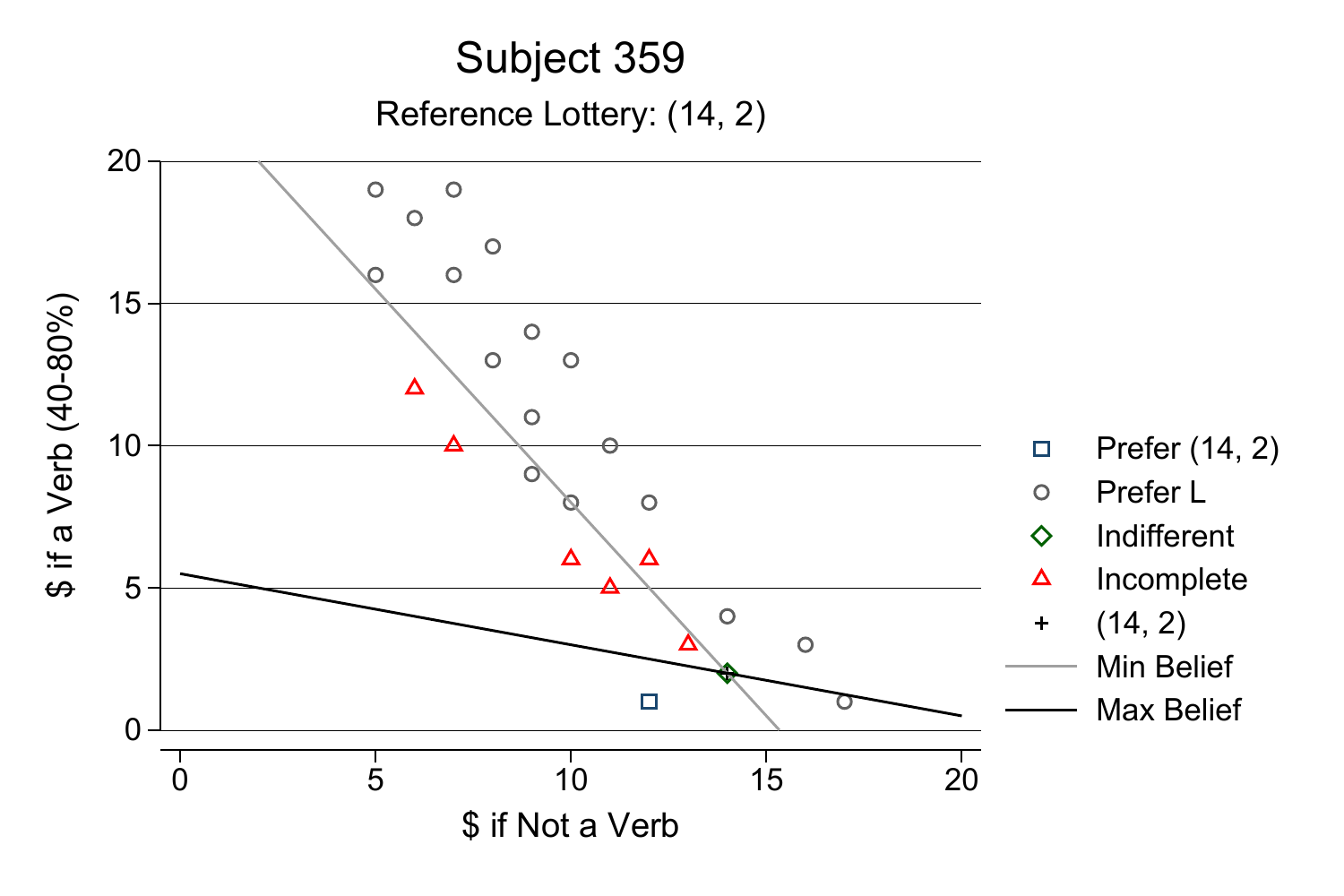}}
    \centerline{\includegraphics[width=0.6\textwidth]{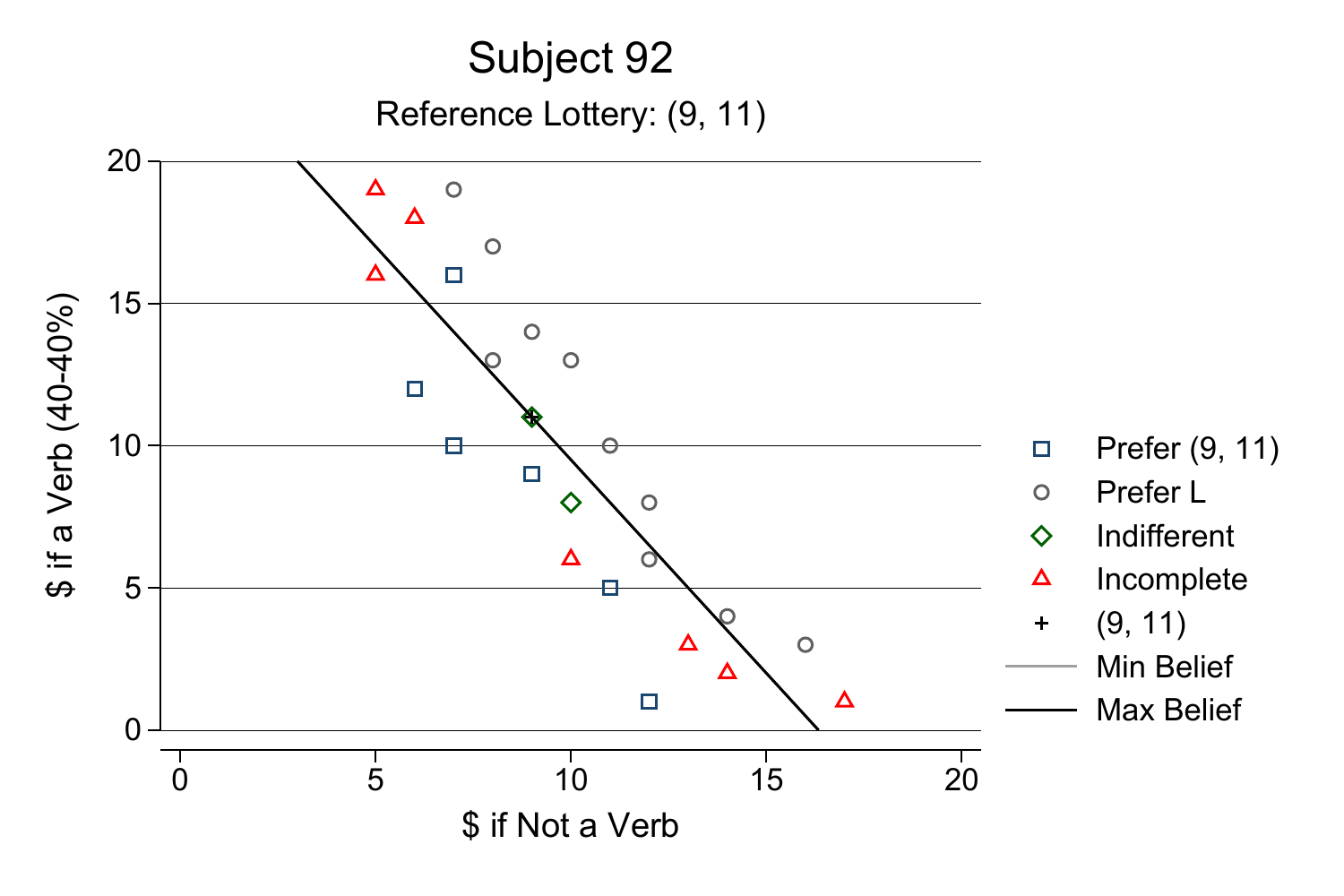}
    \includegraphics[width=0.6\textwidth]{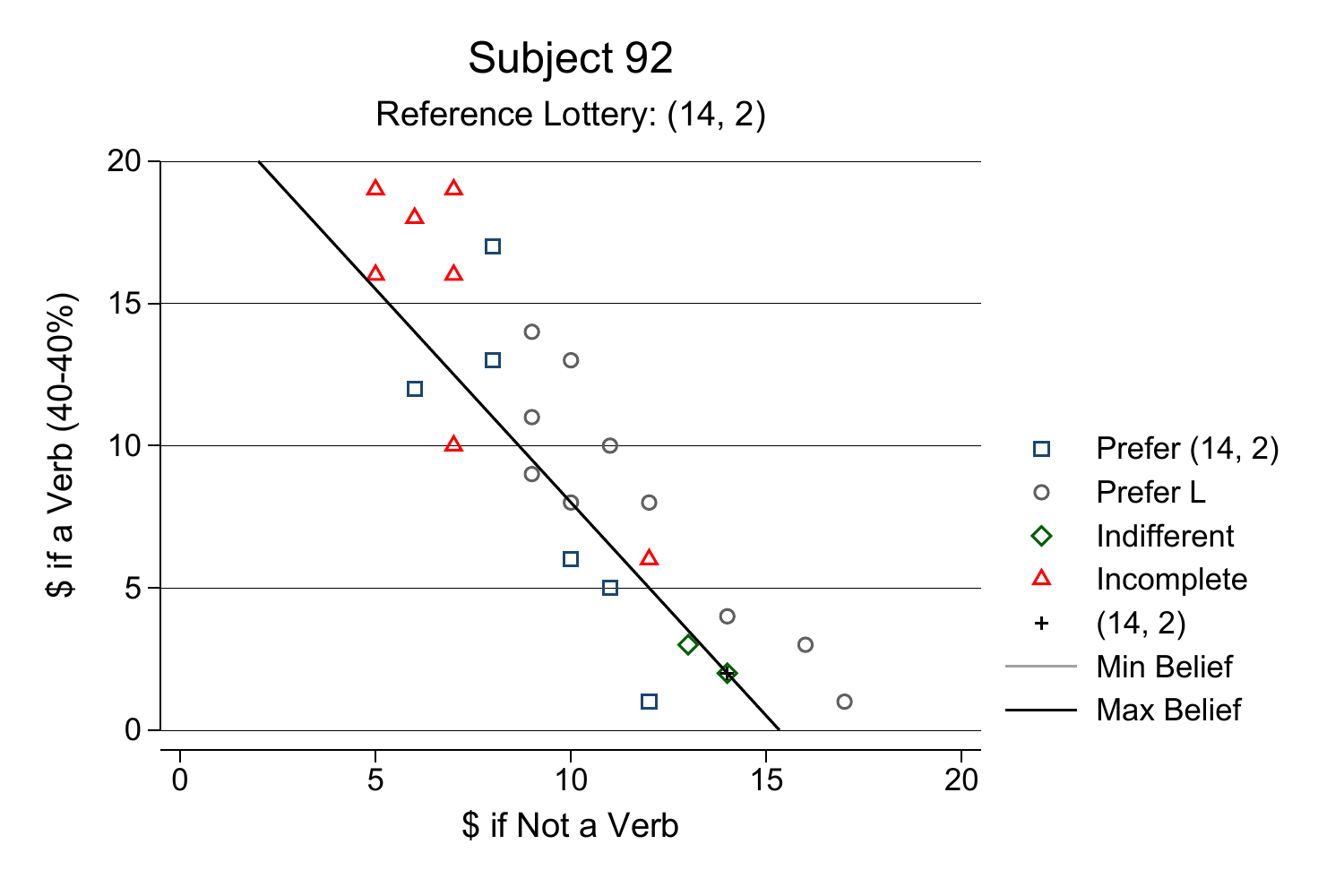}}
    \caption{Three Example Subjects in the Non-Forced Treatment}
    \footnotesize{\underline{Note:} Graphs show the full choice data for three example subjects in the Non-Forced treatment, separated by reference lottery. When subjects reported a range of beliefs, we plot the risk neutral indifference curve implied by the minimum of the range as ``min belief'' and the maximum of the range as ``max belief.'' When subjects reported degenerate beliefs, this belief is reflected in the plot.}
    \label{fig:exampleSubjects}
\end{figure}

Before outlining our main results, we demonstrate the data that we collect using a few representative subjects. Figure~\ref{fig:exampleSubjects} shows the data from the Non-Forced treatment for three subjects; for each subject, we show the two reference lotteries separately. The circles represent lotteries that are preferred to the reference lottery, while the reference lottery is preferred to the comparison lotteries that are marked with squares. Diamonds mark indifference, and triangles mark incompleteness. 

Subject 100---shown in the top two panels---is an example of a perfectly risk-neutral decision-maker who has complete and consistent preferences. They reported a sure belief of $pr(verb)=0.50$, and we plot the risk-neutral indifference curve implied by this belief on each graph. Subject 100 reports indifference for the lotteries that lie on these indifference curves, and reports strict preference for all other lotteries. Lotteries that lie above the indifference curve are preferred to the reference lottery, while the reference lottery is preferred to lotteries lying below the indifference curve. 

In contrast, Subject 359---shown in the middle two panels---has consistent preferences, but they are incomplete. The lotteries that are strictly preferred to the reference lottery (circles) all lie to the northeast of the lotteries that are strictly dis-preferred (squares), emphasizing the consistency of the strict portion of their preferences. However, many lotteries are incomparable to the reference lottery. Subject 359 is not sure of their belief and reported a belief range of $pr(verb) \in [0.4, 0.8]$. Note that their incomplete preferences generally fall within this belief range, suggesting that their incompleteness could result from imprecise beliefs \textit{or} imprecise tastes.

Subject 92---shown in the bottom two panels---is one of our most incomplete subjects, with 14 total incomparabilities. However, Subject 92 reported a sure belief of $pr(verb)=0.40$, so their incompleteness must be due to imprecise tastes.

\subsection{The Prevalence of Incompleteness} \label{sec_mainresults}
\autoref{tab:aggregate_choice_data} reports the aggregate choice data for our two reference lotteries in the Non-Forced treatment. $(9,11)$ is strictly preferred to 55\% of our comparison lotteries, while $(14, 2)$ is strictly preferred to only 15\%. This ensures that our choices sufficiently cover the space of preferences. For both of our reference lotteries, subjects report being indifferent in about 5\% of comparisons, while incompleteness is the least common at 2--3\%. Recall that we had subjects compare the reference lottery to itself; we exclude those comparisons in Table~\ref{tab:aggregate_choice_data}; as we report below, almost all subjects report indifference in these cases.

\begin{table}[!htb]
    \centering
    \begin{tabular}{c|cccc}
    \hline \hline 
    Reference Lottery & Prefer Reference & Prefer Comparison & Indifferent & Incomplete \\ \hline 
     (9, 11)    &  55.3\% & 38.1\% & 4.5\% & 2.1\% \\
     (14, 2)    &  15.4\% & 76.8\% & 4.5\% & 3.3\% \\ 
     \hline \hline 
    \end{tabular}
    \caption{Aggregate Choice Data}
    \footnotesize{\underline{Note:} Subjects made 25 comparisons for each reference lottery. The table presents the percentage of subjects who preferred the reference lottery, preferred the comparison lottery, were indifferent between the two, and were unable to compare the two, aggregated across subjects. Excluded here are the comparisons in which we asked subjects to compare the reference lottery to itself.}
    \label{tab:aggregate_choice_data}
\end{table}

\begin{figure}[!htb]
    \centering
    \includegraphics[width=0.9\textwidth]{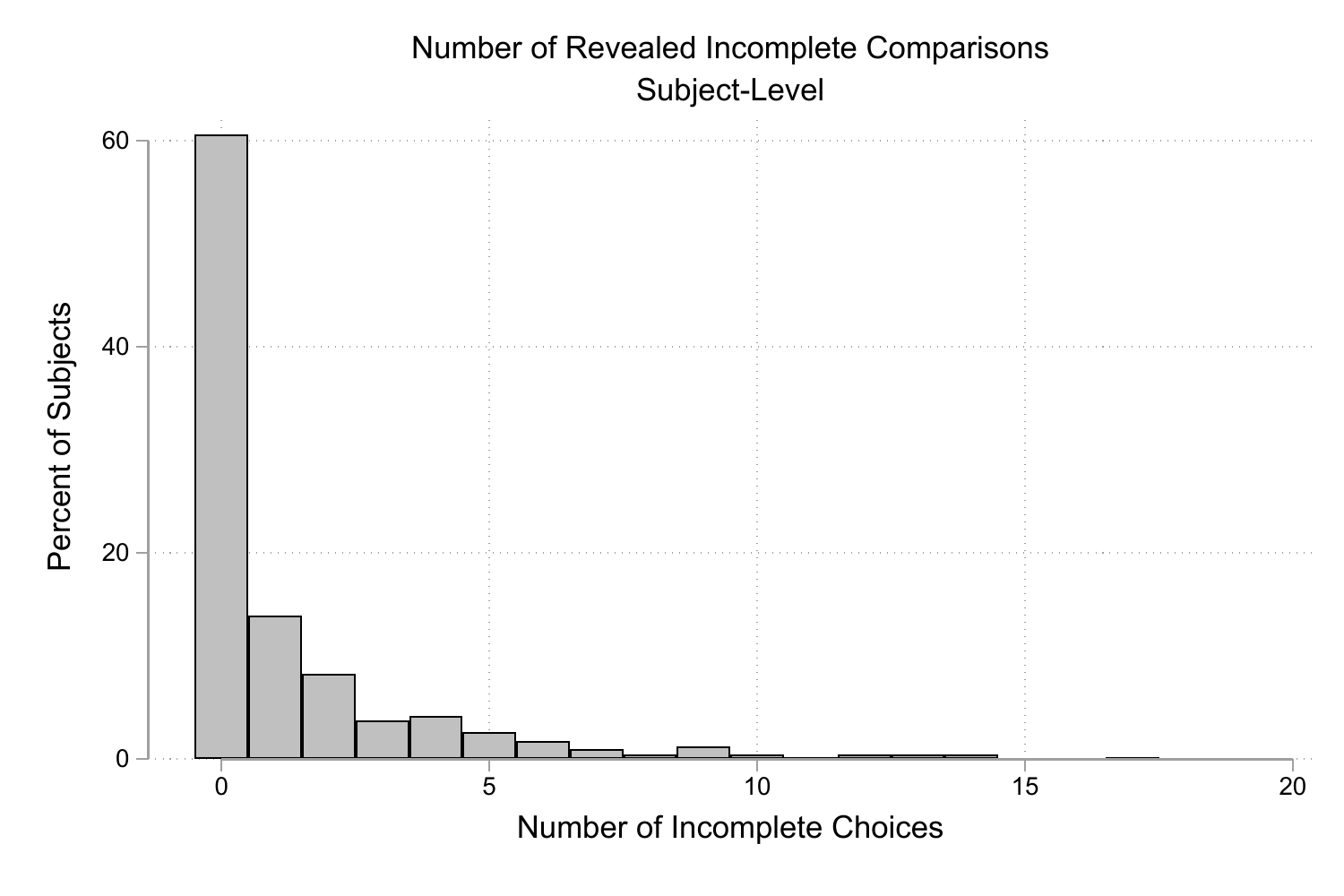}
    \caption{Subject-Level Incompleteness}
    \footnotesize{\underline{Note:} The figure reports the distribution of the number of incomplete comparisons each subject indicated across the two reference lotteries (out of 50 total choices). 61\% of subjects had complete preferences with zero incomplete comparisons.}
    \label{fig:incompleteness_bysubject}
\end{figure}

These averages mask substantial heterogeneity, as evidenced by Figure~\ref{fig:exampleSubjects}. 39\% of subjects (N=252) report some amount of incompleteness, while the remaining 61\% have fully complete preferences in the comparisons that we presented. 
Figure~\ref{fig:incompleteness_bysubject} shows a histogram of the number of directly revealed incomplete comparisons by subject. As the histogram shows, most subjects who report incompleteness do so in relatively few comparisons. Among those with any incompleteness, the average number of incomplete comparisons is 3.3 out of 50 total questions, and the participant with the most incomplete preferences reported 17 out of 50 incomparabilities. 

\paragraph{Demographics} To gain additional insight into this heterogeneity, we look to see the relationship between incompleteness and observed demographics to see if there are any individual correlates with incompleteness. For a subsample of our data, we collected information on subjects' gender, age, education, socioeconomic status, and whether they invested or not.\footnote{We did not collect education and SES in two of our data collection waves, and not all Prolific participants have filled out this demographic information. As a result, we have all demographic variables for 302 subjects.} The only significant predictor of incompleteness is gender---women are significantly more likely to report incompleteness. A full probit regression can be found in Appendix Table~\ref{reg:demographics}.

\begin{figure}[!htb]
    \centering
    \includegraphics[width=0.7\textwidth]{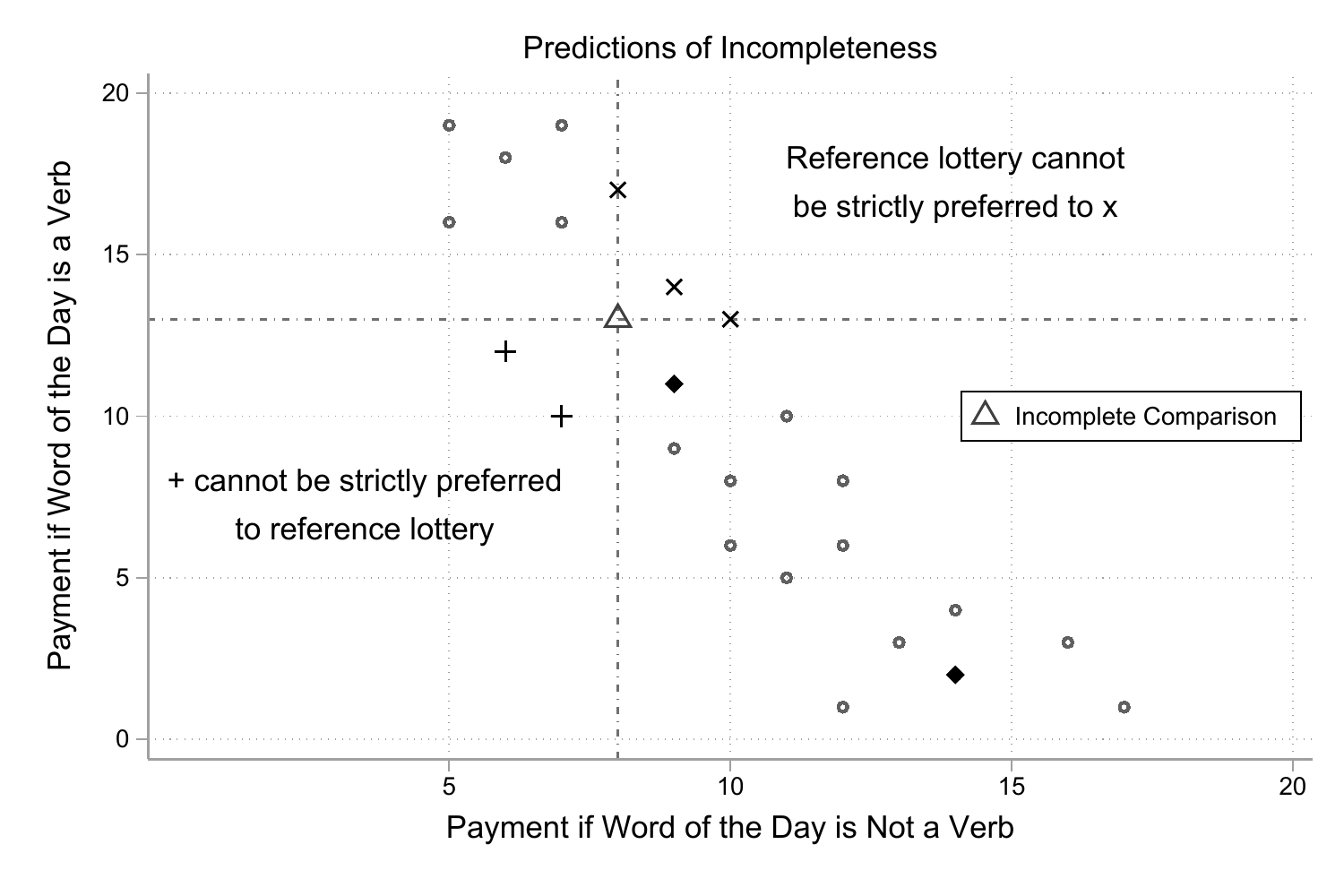}
    \caption{Predictions on Preferences Given Observed Incompleteness}
    \label{fig:predictions}
    \footnotesize{\underline{Note:} Points show the 25 lotteries used in the experiment. Given a reported incomparability---here the example shown is (8,13)---models of incompleteness constrain the reports for dominant and dominated lotteries. The reference lottery should not be strictly preferred to lotteries that dominate (8,13)---lotteries marked with an (x)---and lotteries that are dominated by (8,13)---marked with a (+)---should not be strictly preferred to the reference lottery.}
\end{figure}

\paragraph{Consistency} A natural question is whether individuals report incompleteness in a way that is consistent with theoretical conceptualizations of incomparability. To assess this, we consider a prediction that encompasses both models of imprecise tastes and models of imprecise beliefs. These models predict that if a lottery, $p$, is incomparable to the reference lottery, $r$, then $r$ cannot be strictly preferred to any lottery that dominates $p$ in the sense of paying more in every state. This is because, if $r$ is strictly preferred to the dominating lottery, then it should also be strictly preferred---rather than incomparable---to $p$.\footnote{This is easy to see formally by observing that if $q$ and $r$ are not comparable $E_{\pi }[u(p)] \geq E_{\pi }[u(r)]$ for some $u \in \mathcal{U}$ and $\pi \in \Pi$, and any lottery $q$ that pays more in each state that $p$ will have the property that $E_{\pi }[u(q)] > E_{\pi }[u(p)]$ and therefore $E_{\pi }[u(q)] > E_{\pi }[u(r)]$.} Similarly, any lottery dominated by $p$ cannot be strictly preferred to $r$. Again, this is because, if the dominated lottery were strictly preferred to $r$, then $p$ should also be strictly preferred rather than incomparable to $r$. Essentially, this prediction says that the set of lotteries strictly preferred to the reference lottery must lie above the set of lotteries that are strictly worse than the reference lottery, and incompleteness must lie in between these two sets. See Figure~\ref{fig:predictions} for an illustration. Furthermore, the same prediction should hold for indifference.

We test this prediction by taking every reported incomparability, identifying the lotteries that strictly dominate or are dominated by this lottery, and calculating the percentage of strict preferences for these lotteries that are in the direction predicted, as described above. We find a very high degree of adherence to theory: upwards of $\sim$80\% of strict preferences are consistent with theoretical formulations of incompleteness. We find very similar rates of adherence to the theory for our indifferent comparisons, and we discuss the relationship between incompleteness and indifference more below.

\begin{table}[!htb]
    \centering
    \begin{tabular}{lcc}
    \hline \hline 
    & Incomplete & Indifferent \\ \hline 
    Dominating lotteries preferred to $r$  & 87\% & 89\% \\
    $r$ preferred to dominated lotteries & 78\% & 79\% \\
    \hline \hline 
    \end{tabular}
    \caption{Consistency with Implications of Theoretical Models}
    \footnotesize{\underline{Note:} We calculate this across all incomplete or indifferent comparisons. Percentages report the percentages of strict preferences among dominating/dominated lotteries that are in the direction predicted by theory. The percentages reported consider weak dominance, and the results are even stronger when we consider strict dominance: 87\% and 82\% for incompleteness, and 90\% and 81\% for indifference.}
    \label{tab:consistency}
\end{table}

Our initial results reveal important features of preferences. First, many individuals (about 40\%) have incomplete preferences, even in a very simple stochastic environment. Second, individuals are \textit{aware} of their incompleteness and are able to reveal it directly. Third, this incompleteness behaves systematically and appears in the regions broadly predicted by theoretical formulations of incomplete preferences.   


\subsection{Indifference vs. Incompleteness}\label{sec_indiffvsincomplete}

As evident in Table~\ref{tab:aggregate_choice_data}, subjects are more likely to report indifference than incompleteness. Theoretically, indifference is ``knife-edge'' and should be less prevalent than incompleteness. This brings to question how subjects perceived indifference and incompleteness and how to interpret these choice responses. We present four pieces of evidence suggesting that subjects do differentiate between indifference and incompleteness in a way that aligns with our interpretations.

\paragraph{Survey Evidence} At the end of our study, we asked subjects who ever reported incompleteness why they did so. We gave subjects a few answer options and asked them to select all that applied. 75\% indicated that they ``did not know which gamble (they) preferred,'' and 17\% indicated that they ``did not know how to compare the gambles.'' We also asked subjects who never reported incompleteness to state why they did not use this answer option. 91\% said that they ``always preferred one gamble over the other.'' Less than 1\% of subjects stated that they ``did not trust the algorithm if (they) said (they) didn't know,'' and 1\% said that they ``didn't know what would happen if (they) said (they) didn't know.'' Thus, it does not seem that the low degree of incompleteness is due to distrust or misunderstanding of our incentivization procedure.

\paragraph{Mechanical Indifference and Incompleteness} Second, as a ``sanity check,'' we included each reference lottery as a comparison lottery with itself. That is, subjects were asked to compare $(9,11)$ with itself, and were asked to compare $(14, 2)$ with itself. 93\% of subjects reported indifference in each of these comparisons, and less than 1\% report incompleteness.\footnote{As noted above, these comparisons are excluded from Table~\ref{tab:aggregate_choice_data}, so it does not explain the higher prevalence of indifference than incompleteness.} This gives us reassurance that subjects report indifference when they ``should'' report indifference, and that subjects properly distinguish between indifference and incompleteness. 

The question remains as to whether subjects report incompleteness when they ``should.'' This is harder to detect given that we cannot as easily induce incompleteness like we could with indifference. We attempt to demonstrate this at least directionally with an additional treatment. We recruit 200 new participants through Prolific. For these subjects, we \textit{withhold} payoff-relevant information in some comparisons, and argue that subjects ``should'' be more likely to report incompleteness when they do not have full information. 

In particular, we asked subjects about the following comparisons\footnote{In implementation, we set $x=2$ and $y=5$, since (14, 2) and (5, 19) were lotteries we included in our original treatments.}:

\begin{enumerate}[noitemsep]
    \item $(14, x)$ vs. $(14, x)$
    \item $(14, x)$ vs. $(8, x)$
    \item $(14, x)$ vs. $(14, y)$
    \item $(14, 5-3x+y+(1*-2))$ vs. $(7+(1-x)+2*y-(6+4), 19)$
    \item $(5, 6+5x-2(y+1))$ vs. $(5, 8-y+3x-(2*3))$
\end{enumerate}

We tell subjects that $x$ and $y$ represent some possible payment amount between \$0 and \$20, but we do not tell them the exact amount.\footnote{In implementing the experiment, we used \% and \# symbols rather than $x$ and $y$. We tell subjects that, when the same symbol appears in both options, it represents the same amount of money. When two different symbols appear, they represent different amounts of money.} If subjects understand and trust our algorithm, they would be indifferent between $(14, x)$ and $(14, x)$, despite uncertainty about $x$. Similarly, they would have a strict preference between $(14, x)$ and $(8, x)$. However, they might not be able to compare $(14, x)$ and $(14, y)$, given the uncertainty about $x$ and $y$. It is possible that subjects still form beliefs about $x$ and $y$, enabling them to form a strict preference, or that they are indifferent between them. Nevertheless, we predict that subjects would be more likely to report incompleteness for this comparison than for our comparisons with full information. Finally, we attempt to exaggerate incompleteness by including the last two options that are deliberately complex. We predict that we would see the most incompleteness for these comparisons.

\begin{table}[!htb]
    \centering
    \begin{tabular}{l|ccc}
    \hline \hline 
     & Strict Preference & Indifferent & Incomplete \\ \hline 
     $(\$14, \$x)$ vs. $(\$14, \$x)$  & 5\% & 93\% & 3\%  \\ 
     $(\$14, \$x)$ vs. $(\$8, \$x)$   & 94\% & 4\% & 3\% \\ 
     $(\$14, \$x)$ vs. $(\$14, \$y)$ & 13\% & 51\% & 36\%  \\  
     Complex$_1$ & 46\% & 6\% & 48\%  \\ 
     Complex$_2$ & 39\% & 11\% & 51\% \\
     \hline \hline 
    \end{tabular}
    \caption{Aggregate Choice Data}
    \label{tab:questionmark_data}
    \footnotesize \underline{Notes:} Complex$_1$ refers to $(14, 5-3x+y+(1*-2))$ vs. $(7+(1-x)+2*y-(6+4), 19)$ and Complex$_2$ refers to $(5, 6+5x-2(y+1))$ vs. $(5, 8-y+3x-(2*3))$.
\end{table}

Results, shown in Table~\ref{tab:questionmark_data}, confirm our hypotheses. 93\% of individuals are indifferent between $(14, x)$ and $(14, x)$ and 94\% report strict preference between $(14, x)$ and $(8, x)$. We take this as reassurance that the act of withholding information itself does not lead to incompleteness. However, individuals are much more likely to report incompleteness in comparisons where the lack of information makes it difficult to form a preference: 36\% report incompleteness between $(14, x)$ and $(14, y)$, and half of subjects report incompleteness in our complex comparisons. While we still cannot say whether the remaining half of subjects ``should'' report incompleteness here, results demonstrate, at least directionally, that individuals are willing to report incompleteness when they do not know how to compare two alternatives. Indeed, 76\% of individuals report incompleteness at least once in this treatment, compared to 39\% in our original data (Fisher's exact, $p<0.001$). 

\paragraph{Location} As an additional comparison of incompleteness and indifference, we find that subjects report indifference for very different lotteries than those for which they report incompleteness, again suggesting that individuals treat these response types differently. Individuals tend to report incompleteness for lotteries that are ``far'' from the reference lottery, while they report indifference for lotteries closer to the reference lottery; see Figure~\ref{fig:exampleSubjects} for examples. We quantify this using the Euclidean distance from the reference lottery to the comparison lottery. Table~\ref{tab:euclideandistance} reports the average distance from the reference lottery for indifferent and incomplete comparisons.\footnote{We exclude the comparisons where individuals compare the reference lottery to itself, as these will mechanically reduce the Euclidean distance for indifferent comparisons.} The average distance is significantly higher for incomplete responses than indifferent responses. Our $(14,2)$ reference lottery emphasizes this result, since this lottery is very skewed and many comparison lotteries are relatively far from it in Euclidean space. This also explains why we see more responses indicating incompleteness in this treatment in aggregate.

\begin{table}[!htb]
    \centering
    \begin{tabular}{c|ccc}
    \hline \hline 
    &  & & ranksum \\
    Reference Lottery & Indifferent & Incomplete &  $p-$value \\ \hline 
     (9, 11)    &  4.72 & 6.32 & $<0.0001$ \\
     (14, 2)    &  5.96 & 9.20 & $<0.0001$ \\
     \hline \hline 
    \end{tabular}
    \caption{Average Euclidean Distance from Reference Lottery}
    \label{tab:euclideandistance}
    \footnotesize{\underline{Note:} For comparisons where subjects indicated indifference or incompleteness, we calculate the Euclidean distance between the reference lottery and comparison lottery. The table presents the average distance across subjects for both of these response options.}
\end{table}

We find this result interesting for a few reasons. First, it gives a unique prediction for models of incompleteness to capture. Second, it suggests that incompleteness is exacerbated for comparisons that involve objects which are very different from one another. More work is needed to understand this, but it suggests that local trade-offs are easier to make than large trade-offs across dimensions.

\paragraph{Response Times} Finally, we analyze response times conditional on choice. In each comparison, we record the time it takes a subject to submit their decision. Subjects take 8.84 seconds to submit a strict preference, on average. This is directionally though not significantly faster than indifference (9.73 seconds, $p=0.147$) and incompleteness (10.49 seconds, $p=0.076$).\footnote{We exclude exact comparisons in this calculation, since the response time literature typically does not ask subjects to choose between two identical objects. $p-$values are reported from a linear regression with standard errors clustered at the subject level.} Thus, it appears that incompleteness is the slowest response type in our data.

We confirm this relationship in the Forced treatment, as well. We identify comparisons where subjects reported indifference or incompleteness in the \textit{Non-Forced} treatment, and look to see their response times in these comparisons when we force them to choose. Again, we find that subjects take longer in decisions where they are being forced to complete their preference, though the differences are not statistically significant (strict preference: 7.59 sec; incompleteness: 7.68 sec; $p=0.867$).

Taken all together, we find evidence that individuals do have incompleteness in their preferences, and they treat this separate from indifference.

\begin{result}
Over one third ($\approx$ 40\%) of subjects directly reveal incompleteness in their preferences. Secondary evidence agrees with an interpretation of these incomparabilities as evidence of incompleteness. However, while we can detect incompleteness in preferences, it is relatively rare and appears in only a few comparisons, on average, for a given subject.
\end{result}

\subsection{The Source of Incompleteness}\label{sec_beliefsvstastes}
Given that we observe incompleteness in preferences over uncertainty, a natural question is whether this incompleteness results from individuals' tastes or their beliefs. One theory is that incompleteness reflects imprecision in \textit{beliefs} \citep{bewley2002knightian}, while an alternative hypothesis is that incompleteness reflects imprecision in \textit{tastes} \citep{aumann1962completeness}. We look to see whether subjects with incomplete preferences are more likely to have imprecise beliefs, lending support for the first hypothesis, or instead whether incompleteness is as prevalent in individuals with precise beliefs, indicating a source of imprecise tastes.

\begin{adjustwidth}{-100pt}{-100pt}
\begin{figure}[!htb]
    \centering
    \includegraphics[width=0.49\textwidth]{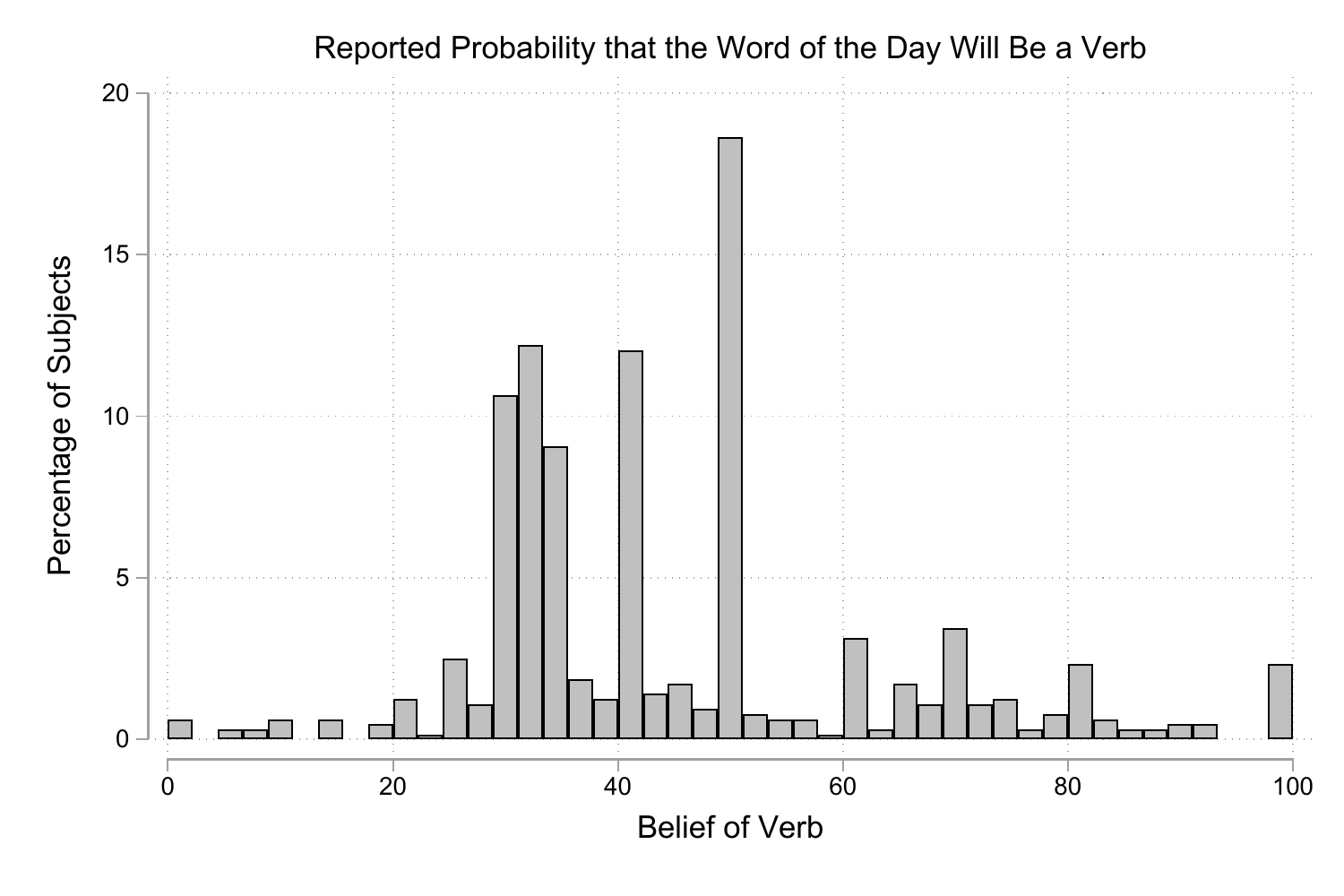}
    \includegraphics[width=0.49\textwidth]{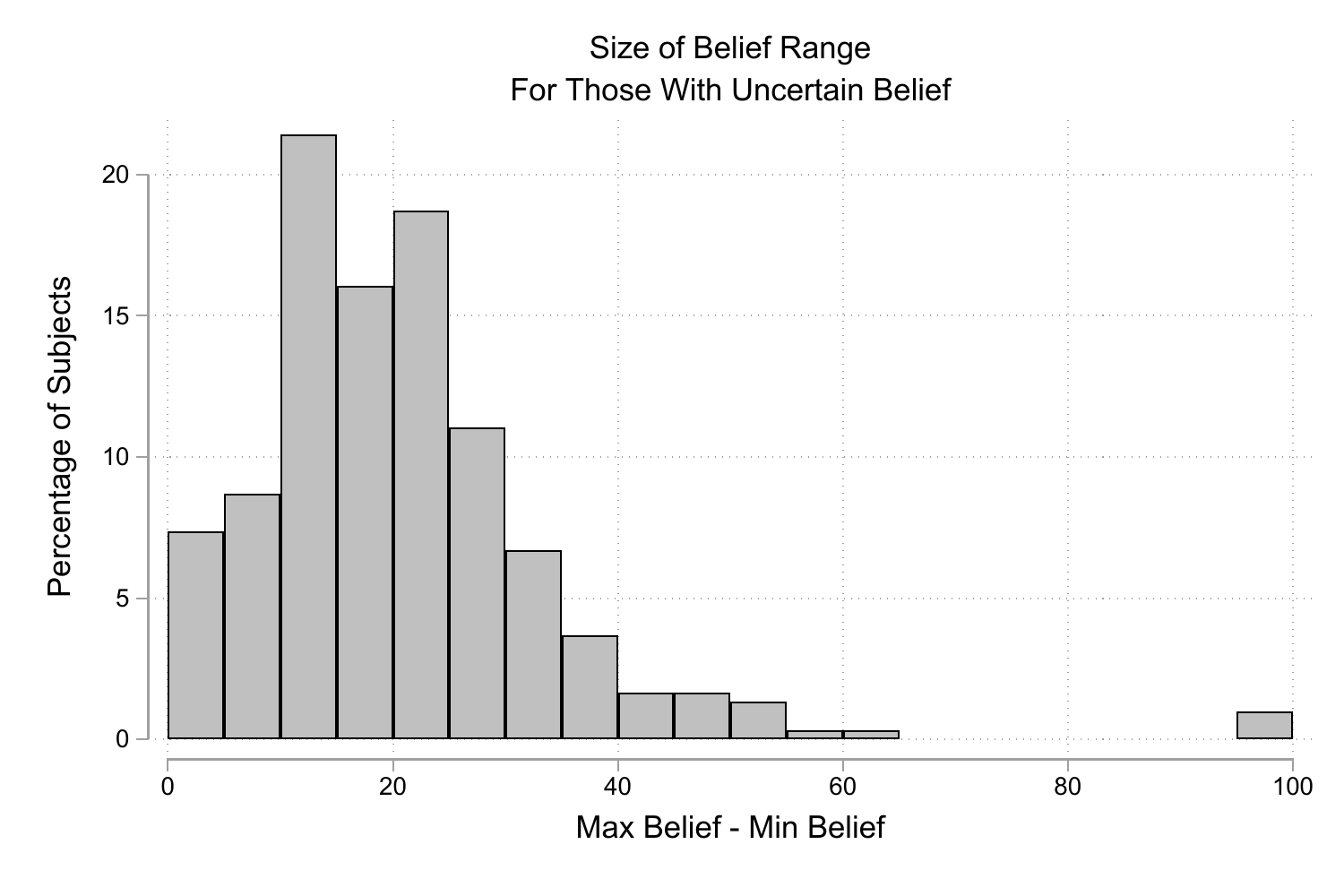}
    \caption{Distribution of Reported Beliefs of the Likelihood that the Word of the Day will be a Verb}
    \footnotesize{\underline{Note:} The left panel presents the distribution of reported beliefs that the word of the day would be a verb. The right panel presents the distribution of the size of belief ranges for the 47\% of subjects who reported having an uncertain belief.}
    \label{fig:beliefverb_histogram}
\end{figure}
\end{adjustwidth}

The left panel of \autoref{fig:beliefverb_histogram} shows the distribution of reported beliefs that the Word of the Day will be a verb. Most subjects are clustered around a belief of one-third, with another group clustered around one-half. 47\% of subjects indicate uncertainty about their reported belief. This is on par with the 49\% of subjects who report uncertainty about developing late-onset dementia---a dramatically different event---in \citet{giustinelli2019precise}. Subjects who report uncertainty about their belief have the opportunity to report a range of beliefs, and on average, they report ranges that span 19 percentage points. The full distribution of belief ranges for those with uncertain beliefs can be found in the right panel of Figure~\ref{fig:beliefverb_histogram}.

If incompleteness stems mainly from uncertainty in beliefs, we would expect the subjects who report incompleteness to be the subjects who indicate uncertainty about their beliefs. Table~\ref{tab:incomplete_byBelief} shows the joint distribution between incompleteness and belief uncertainty. Of those with incomplete preferences, only half have uncertain beliefs. Among subjects with belief uncertainty, 43\% report incomplete preferences, directionally but only marginally significantly higher than the 36\% reporting incompleteness among those with certain beliefs (Fisher's exact, $p=0.075$). Furthermore, among the subsample of subjects with imprecise beliefs, we find no significant correlation between tendency to report incompleteness and the size of the belief range (correlation: 0.0447, $p=0.441$). Thus, despite evidence that many subjects have imprecise beliefs, we do not find evidence that this is the main contributor to incompleteness in preferences.

\begin{table}[!htb]
    \centering
    \begin{tabular}{lcc}
    \hline \hline 
        & \multicolumn{2}{c}{Preferences} \\
        & Complete & Incomplete  \\ \hline 
    Certain Belief (53\% of subjects)     & 34.0\% & 19.2\% \\
    Uncertain Belief (47\% of subjects) & 26.6\% & 20.2\% \\ \hline \hline 
    \end{tabular}
    \caption{Relationship Between Incompleteness and Imprecise Beliefs}
    \label{tab:incomplete_byBelief}
    \footnotesize{\underline{Note:} The table presents the aggregate percentage of subjects broken down by whether they have certain or uncertain beliefs and whether they have complete or incomplete preferences. Reported percentages are unconditional.}
\end{table}

\paragraph{Objective Lotteries} To confirm the previous finding, we run two additional treatments, each with about 150 new subjects recruited through Prolific. These treatments are exactly the same as our original experiment, except that we provide subjects with an objective probability. Specifically, in one treatment, we tell subjects that there is a $\frac{1}{3}$ chance of the event realizing---analogous to a $\frac{1}{3}$ chance that the word of the day would be a verb---and in the other, we tell subjects that there is a $\frac{1}{2}$ chance of the event realizing.\footnote{Since we want to truthfully tell subjects the objective probabilities while keeping everything as similar as possible to our subjective treatments, we show subjects pictures of cards that have ``verb'' or ``not a verb'' written on them. For example, in the treatment where we induce an objective prior of $\frac{1}{3}$, we show subjects three cards, one of which says ``verb'' and the other two say ``not a verb.'' We tell subjects that we will randomly select one of these cards and the label of the card will determine their payment from a given lottery. We tell subjects that they can watch us live-streaming the card draw on Twitch in three days to further instill trust in the randomization process. We did live-stream the card draws, but did not have any viewers.} These objective probabilities reflect the modal beliefs from our subjective version of the experiment, allowing us to make comparisons about the likelihood of incompleteness across objective and subjective uncertainty. We still elicit subjects' beliefs at the end of the experiment and we ask them whether they are certain about this belief. 

First, we find that individuals are indeed more sure of their beliefs in our objective treatments. Aggregating across all subjects in both objective treatments, only 11\% indicated that they were unsure about their reported belief. This is significantly lower than the 47\% who indicated belief uncertainty in the subjective treatment (Fisher's exact, $p<0.001$).\footnote{Of those who were sure of their belief, 78\% and 88\% of subjects reported a belief equal to the objectively probability in the $\frac{1}{3}$ and $\frac{1}{2}$ treatments, respectively.} This further confirms that our subjective event---the Merriam Webster Dictionary Word of the Day---induced true subjective uncertainty over which individuals were unable to form precise beliefs. In contrast, they were sure of their beliefs about our objectively-given probabilities.

At an individual level, we find no significant reduction in the tendency to report incompleteness when lotteries are defined over events with objective probabilities. Recall that, in our original data, 39\% of subjects reported incompleteness at least once. We find a similar result in our new objective treatments: 37\% of subjects report incompleteness at least once when probabilities are objectively given (Fisher's exact, $p=0.615$). 

To look at incompleteness on the level of an individual question, we need to compare individuals who have similar beliefs, since the predicted extent of incompleteness depends on one's belief. Furthermore, our main goal is to analyze the extent to which belief uncertainty contributes to incompleteness, so we want to consider individuals with similar beliefs but who are either certain or uncertain about this belief. Motivated by this, we compare individuals with beliefs near one-third or one-half in the subjective treatment who are \textit{uncertain} about this belief to individuals with the same beliefs in the objective treatment who are \textit{certain} about their belief. The number of subjects who fit these criteria are referenced in Table~\ref{tab:aggregate_choice_data_objective}. For this analysis, we call reported beliefs in $[30, 35]$ ``near one-third'' and beliefs in $[48, 52]$ ``near one-half'' in order to allow for small deviations in reporting.

\begin{table}[!htb]
    \centering
    \begin{tabular}{cc|ccc}
    \hline \hline 
                           & & Strict Preference & Indifferent & Incomplete \\ \hline 
    \multirow{2}{*}{Near One-Third} & Uncertain (N=92)  &  93.3\% & 3.8\% & 2.9\% \\
     & Certain (N=117) &  93.5\% & 3.4\% & 3.1\% \\ \hline 
    \multirow{2}{*}{Near One-Half} & Uncertain (N=50) &  91.9\% & 5.1\% & 3.0\%  \\
     & Certain (N=126) &  94.5\% & 3.5\% & 2.0\% \\ 
     \hline \hline 
    \end{tabular}
    \caption{Aggregate Choice Data}
    \footnotesize{\underline{Note:} The table presents the percentage of subjects who had a strict preference, were indifferent, and were unable to compare the gambles, aggregated across subjects and reference lotteries. Excluded here are the comparisons in which we asked subjects to compare the reference lottery to itself. Uncertain $\frac{1}{3}$ are subjects who reported beliefs in $[30, 35]$ in the subjective treatment and reported that they were not sure of this belief; Certain $\frac{1}{3}$ are subjects who reported beliefs in $[30, 35]$ in the objective treatment and reported that they were sure of this belief. The $\frac{1}{2}$ are defined analogously in the range $[48, 52]$. Sample sizes, N, are reported as number of subjects who satisfy the given belief restrictions. }
    \label{tab:aggregate_choice_data_objective}
\end{table}

Table~\ref{tab:aggregate_choice_data_objective} presents the aggregate choice data. For beliefs near one-third, we find extremely similar rates of incompleteness. For beliefs near one-half, we find slightly more incompleteness reported by individuals with uncertain beliefs in our subjective treatment compared to individuals with certain beliefs in our objective treatment (rank-sum $p=0.485$ for beliefs near one-third, $p=0.0066$ for beliefs near one-half). Nevertheless, most of the incompleteness remains even with objective probabilities that individuals are certain about. Thus, we conclude that belief uncertainty does not play a large role in contributing to incompleteness. 

\begin{result}
Imprecise beliefs do not fully explain incompleteness. Of those with incomplete preferences, only half report being uncertain of their subjective belief. Furthermore, individuals are equally likely to express incompleteness in objective and subjective environments.
\end{result}

\subsection{The Impact of Forced Choice}\label{sec_forcedchoice}
Given that typical choice experiments follow the structure of our Forced Choice design (eliciting preferences without allowing for explicit reports of indifference or incompleteness), we compare choices in the Non-Forced treatment to those in the Forced treatment to see how indifference and incompleteness affect inferences made from forced choice designs. The idea is similar to \citet{mandler2005incomplete} and \citet{nishimura2016incomplete}. They model two preference relations: \textit{choices} are complete by construction, but might be intransitive, while \textit{tastes} can be incomplete, but the complete portion of the relation is transitive. Under this interpretation, our Non-Forced treatment measures tastes and the Forced treatment measures choices. We analyze the extent to which these preference relations agree, and the extent to which intransitive choices can be explained by incomplete tastes.

First, we compare choices in the Forced and Non-Forced treatments \textit{where subjects reported a strict preference in the Non-Forced treatment}. Here, subjects were given the option to report indifference or incompleteness and did not, so we would predict that they should report the same preference in the Forced treatment. Indeed, subjects give the same response in 86\% of such instances. This is a high degree of consistency in choices where subjects have a strict preference.

Despite the high degree of consistency, we find that subjects exhibit preference reversals in 14\% of choices in which they revealed a strict preference. Although this could be evidence of stochastic preferences or ``learning'' one's preferences throughout the experiment, another interpretation is that these preference reversals reflect underlying ``indirectly revealed'' incompleteness that subjects themselves are not aware of \citep{bayrak2017reversals}. We find strong suggestive evidence of this interpretation by comparing the percentage of subjects who report incompleteness in a given question to the percentage of subjects who exhibit a preference reversal in the same question. That is, for each of our comparison lotteries, we calculate the percentage of subjects who report incompleteness when comparing this lottery against one of the reference lotteries in the Non-Forced treatment, and we correlate this with the percentage of subjects who report a strict preference in the Non-Forced treatment but then choose the other lottery in the Forced treatment. Figure~\ref{fig:reversals} presents these results separated by reference lottery.

\begin{adjustwidth}{-100pt}{-100pt}
\begin{figure}[!htb]
    \centering
    \includegraphics[width=0.49\textwidth]{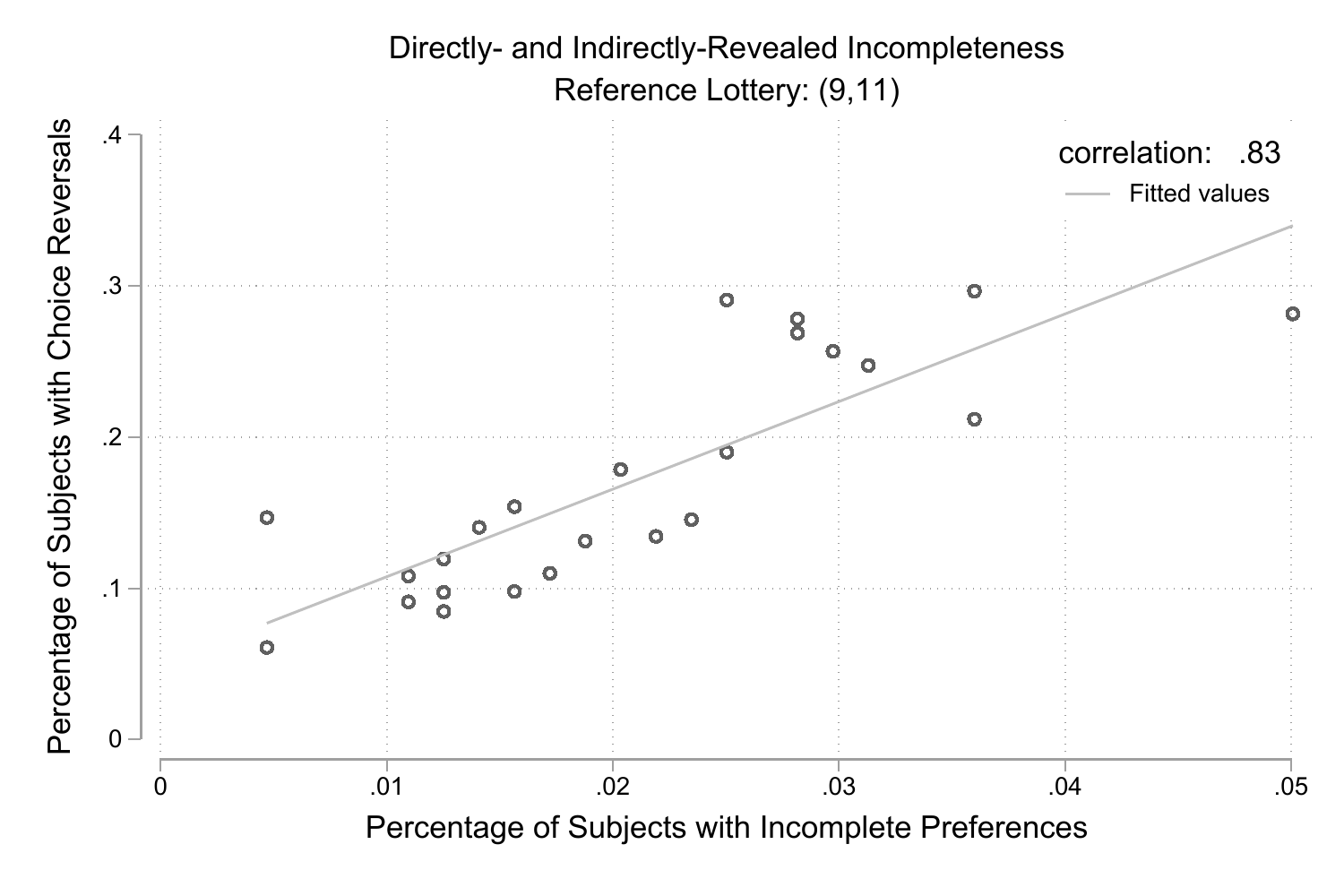}
    \includegraphics[width=0.49\textwidth]{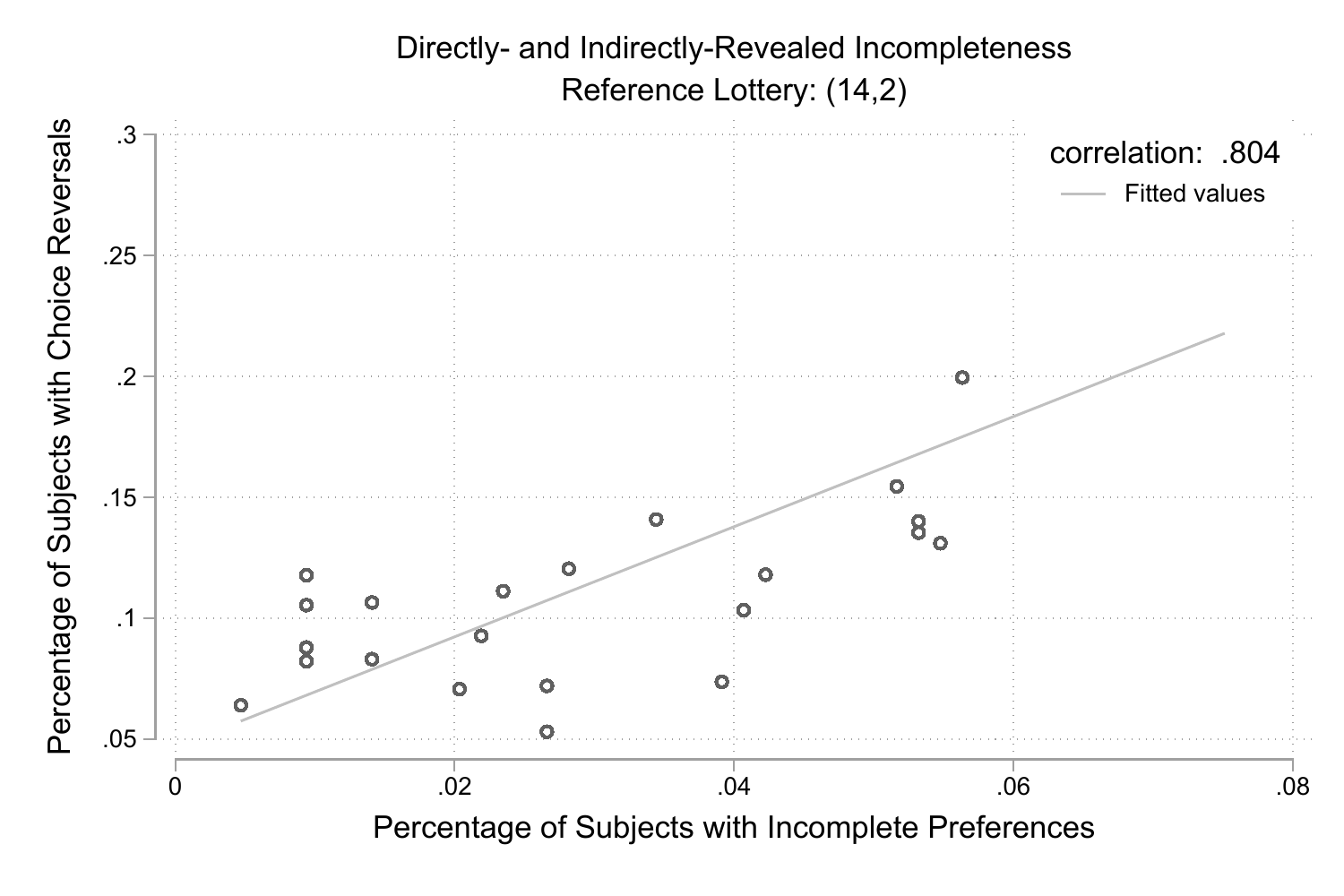}
    \caption{Preference Reversals as a Potential Indicator of Underlying Incompleteness}
    \footnotesize{\underline{Note:} Each point on a graph represents one of our 24 comparison lotteries; we exclude questions in which subjects compare the reference lottery to itself. The two panels separate the two reference lotteries.}
    \label{fig:reversals}
\end{figure}
\end{adjustwidth}

We find a strong positive correlation between these two percentages: The comparisons in which subjects are more likely to report incompleteness are the same comparisons in which other subjects are likely to exhibit preference reversals (correlation of 0.83 for reference lottery (9,11) and 0.80 for reference lottery (14,2)). This provides suggestive evidence in favor of the interpretation that preference reversals can be understood as indirectly revealed incompleteness---when we include both directly and indirectly revealed incompleteness, 98\% of subjects have incomplete preferences. Furthermore, this provides evidence that there may be comparison-specific features of decisions that make it more difficult to form preferences. 

Finally, we note that almost all subjects (94\%) who directly reveal incompleteness in the Non-Forced treatment also indirectly reveal incompleteness by exhibiting a preference reversal. Thus, we do not interpret these preference reversals to mean that subjects were distrustful of our experiment and were unwilling to reveal incompleteness directly, since these preference reversals occur at similar rates in subjects who we know were willing to reveal incompleteness directly. Instead, this suggests that even individuals who are sometimes aware of their incompleteness have underlying incompleteness that they are not aware of. 

\begin{result}
The percentage of subjects who report incompleteness in a given question is strongly and positively correlated with the percentage of subjects who exhibit a preference reversal in the same question, suggesting that preference reversals reflect indirectly revealed incompleteness. If we include these preference reversals as incompleteness, 98\% of subjects have incomplete preferences.
\end{result}

Next, we compare intransitivities across Forced and Non-Forced choices. Our experimental design allows us to test transitivity as follows. In both the Forced and Non-Forced blocks, individuals compare lotteries, $p$, to our two reference lotteries, $r_1$ and $r_2$. As one of these comparisons in each block, individuals also compare $r_1$ and $r_2$ directly. An individual's ranking between $r_1$ and $r_2$ allows us to ``link'' all of the choices together by transitivity. For example, an individual who prefers $p$ to $r_1$, and who prefers $r_1$ to $r_2$, should also prefer $p$ to $r_2$. As such, a violation of transitivity is observed whenever $p \succeq  r_i \succeq r_j \succeq p$.\footnote{Recall that we had subjects compare the two reference lotteries to each other twice. It is possible that a subject reports a different preference across these two choices. We consider it a violation of transitivity if either report constitutes a violation. }

In our Forced choice treatment, individuals are only given the option to report $p \succeq r_i$ or $r_i \succeq p$, so the expression above presents the only possible transitivity violations. In our Non-Forced treatment, individuals can report indifference explicitly, which presents additional transitivity violations. For example, $p \sim r_i \succ r_j \succ p$ represents a Non-Forced transitivity violation, as well. We analyze Non-Forced violations first only in strict preference and then include indifference.

In forced choice, we find 4.0\% of choices constitute a transitivity violation. This is significantly higher than the 1.7\% of strict preferences in the Non-Forced treatment that constitute transitivity violations ($p<0.0001$). This gap shrinks but is still significant when we included Non-Forced intransitivities that involve indifference (3.3\%, $p=0.0184$). This is stark, since individuals have four answer options in the Non-Forced treatment so trembles would be more likely to lead to intransitivities. Of the transitivity violations observed in Forced choice, 29\% involve a comparison identified as indifferent or incomplete in the Non-Forced treatment. This is significantly higher than the percentage of transitive comparisons that include an indifferent or incomplete comparison (22\% vs. 29\%, Fisher's exact $p<0.001$). Thus, a significant proportion of intransitivities in Forced choice can be explained by incompleteness or indifference.

\begin{result}
Forced choice leads to more inconsistencies in preferences compared to Non-Forced choice. Incompleteness can explain a significant portion of intransitivities in choice.
\end{result}

Taking our evidence altogether, we find that a significant portion of individuals do have incomplete preferences in a simple stochastic environment. This incompleteness appears in predictable areas and individuals are aware of a significant portion of incompleteness in their preferences. Allowing individuals to tell us when they don't know their preference leads to more coherent elicited choices. Most of the incompleteness we find seem to be attributable to imprecise tastes rather than imprecise beliefs.

\subsection{Did Subjects Report Truthfully?}\label{sec:manipulation}
Naturally, one might worry about how subjects perceived our elicitation and whether they reported truthfully. Our design allows for many ways in which to evaluate subjects' responses and detect evidence of manipulation. We collect the evidence here, and believe that it paints a picture of truthful reporting. 

First, we look for evidence to evaluate subjects' reports of strict preferences. There are two features of the data that suggest subjects reported strict preferences truthfully. As a first piece of evidence, we find that subjects who report a strict preference when incentivized by our mechanism report the same choice 86\% of the time under standard incentives, a high degree of consistency. As a second piece of evidence, we included a few lottery comparisons that were related by strict dominance (i.e., one lottery paid strictly more in both states). In these questions, subjects report to our mechanism a strict preference for the dominant lottery in 89\% of instances. Taken together, our data suggest that subjects were not reporting ``untruthful'' strict preferences when incentivized by our mechanism. 

Second, we look for evidence to evaluate subjects' reports of indifference. Our main validation of these reports is when we ask subjects to compare a lottery to itself. Here again, over 90\% of subjects report indifference. Thus, we find compelling evidence that subjects do not attempt to manipulate our mechanism---by reporting a strict preference or incompleteness---when they are actually indifferent. 

Given that we have evidence that subjects report strict preferences and indifferences truthfully, we find it natural to conclude that subjects also report incompleteness truthfully. This is the hardest to validate in the data, precisely for the fact that incompleteness is hard to identify. Nevertheless, there are three features of the data to support this claim. First, as reported in Section~\ref{sec_mainresults}, we find that $\sim$80\% of reported incomparabilities are consistent with theoretical conceptualizations of incompleteness. Second, we find that over twice as many subjects report incompleteness (from 39\% of subjects to 76\% of subjects) in our treatment that deliberately increases complexity and removes information, which was designed to increase rates of incomparability. Third, we see a strong positive correlation (across-subject) between incompleteness that is revealed directly through our mechanism and incompleteness that is revealed indirectly through preference reversals. If one believes preference reversals and stochastic choice to be an indicator of incompleteness, then our mechanism measures behavior that is highly correlated with this. Finally, we note that survey responses confirm our interpretation, as well: 91\% of subjects who do not report incompleteness say that it is because they always preferred one over the other, and 84\% of subjects who report incompleteness say that it's because they did not know which gamble they preferred or did not know how to make the comparison between the two gambles.

Taken together, we believe there is compelling evidence that our experimental participants reported their preferences thoughtfully and truthfully. This suggests that estimating preferences and paying based on estimated preferences is \textit{behaviorally} incentive compatible \citep{danz2022belief}. As discussed in Appendix~\ref{sec:appendix_algorithms}, there are many ways in which we could change the precise way in which we carried out this estimation in order to address different potential sources of manipulation. We chose to start with the simplest ``estimation'' as a proof-of-concept, and our data suggest that this procedure elicited valid and reliable information on preferences.

\section{Related Literature} \label{sec:literature}
Our paper is most closely related to other papers that have attempted to identify incompleteness in preferences (see \citealp{bayrak2020imprecision} for a survey of the literature and methodologies related to preference imprecision). We overview these related methodologies and how our results compare to what has been found in the literature. We note, however, that we are not aware of any papers that exist in an environment directly comparable to ours, so we cannot directly say whether the prevalence and shape of incompleteness is comparable to the incompleteness found in other methodologies. 

\citet{cohen-jaffray-said_85} and \citet{cohen-jaffray-said_87} are the first experimental papers we are aware of in which subjects are presented with an ``indifference'' and an ``I do not know'' option. In their case, subjects compare a certainty equivalent to a binary risky option with given objective probabilities using a price list (they also have an unknown probabilities option). They find that about 10\% of subjects display indecision. In their case, reporting ``I do not know''  implied the experimenter chose for the subject.

\citet{cubitt2015imprecision} have subjects fill out a standard multiple price list by reporting a switch point, but also allow subjects to report that they are ``not sure about (their) preference'' in any rows. However, in order to incentivize these decisions, subjects were also required to complete their incomplete preferences by indicating a single switch point even if they were unsure. Across all of their questions, they find that 87\% of subjects report some preference imprecision by using this ``unsure'' option. 

This is in line with \citet{agranov2020ranges} who ask subjects to fill out a multiple price list but allow for the option to explicitly randomize in each row. For subjects who choose to randomize across multiple consecutive rows, one interpretation is that preferences are imprecise in this range. They find that between 50--75\% of subjects choose to randomize in a way that is consistent with preference imprecision. This is also consistent with their earlier work in which 70\% of subjects randomize across multiple repetitions of the same decision \citep{agranov2017randomization}.

While one interpretation of randomization is that subjects have imprecise preferences, individuals might randomize instead if they have convex preferences, utility from gambling, misspecified beliefs about the uncertainty generating process, or other potential explanations \citep{agranov2021stable}. Thus, the literature has attempted to identify evidence of incompleteness in other ways. In a related paper, \citet{costagomes2021deferral} use costly deferral as indication of incompleteness. Subjects are presented with choices over consumption goods (in their case, headsets). In one treatment, subjects were forced to make a choice of headset, while in the other they could pay a small cost to defer choice to the end of the experiment. They find that 35\% of subjects use the deferral option. They also find that choices are more coherent in the deferral treatment than in the forced choice treatment, similar to our results comparing the Forced and Non-Forced treatments.

These experiments used very different elicitation methodologies, but our 39\% of subjects expressing incompleteness is in line with other estimates in the literature. Randomization seems to be more prevalent than deferral or explicit statements of incompleteness. This could be because randomization captures additional preferences as discussed above, or it could be that deferral and explicit incompleteness require more ``sophistication'' and awareness of incompleteness. This is consistent with the conclusion from \citet{cettolin2019incomplete}, which is that some amount of randomization can be attributed to incompleteness while for other subjects it is a manifestation of their inherent preference for randomization. 

Our paper builds on this previous literature and contributes to the understanding of incomplete preferences in a few ways. First, our elicitation procedure reveals when individuals ``do not know'' how to make a choice or are unsure of their preference. This methodology is designed to capture welfare-relevant incompleteness, where individuals do not want the analyst to make inference based on these decisions. We view this as complementary to other elicitation methodologies in the literature and think it would be interesting to compare what is revealed across these different methodologies. Second, we design our experiment to conform to theoretical studies of incompleteness, and directly elicit from subjects a distinction between indifference and incompleteness. Third, we separate imprecise beliefs from imprecise tastes, and find that incompleteness results from imprecise tastes more than imprecise beliefs. Fourth, and related to the previous point, we study incompleteness in a domain of subjective uncertainty while many previous papers have focused on objective uncertainty. We use a novel but natural event structure that allows for imprecise subjective beliefs, but are still able to compare this to an equivalent environment with objective uncertainty. Finally, we confirm what \citet{costagomes2021deferral} find---that forced choice results in less consistent decisions---and our experiment further suggests that preference reversals indicate underlying incompleteness.

Finally, from a methodological perspective, using a preference estimation algorithm to elicit otherwise hard-to-elicit information is very much in the spirit of \citet{krajbich2017neurometrically} and \citet{kessler2019resume}, though neither of these papers use the mechanism to allow for incompleteness. \citet{krajbich2017neurometrically} use an adaptive algorithm to estimate subjects' preferences in a prospect theory model. To get a better estimate of preferences, their experiment dynamically chooses which questions to ask subjects based on their earlier responses. An issue with this is that subjects could distort their earlier decisions in order to manipulate the questions they see later. To restore incentive compatibility, the authors tell subjects that the choices they face are hypothetical, but that these choices will be used to estimate their risk preferences which will determine their payment. 

\citet{kessler2019resume} conduct a similar exercise to detect discrimination without running a resume audit study. They had employers rate resumes known to be hypothetical. They used these ratings to estimate employers' preferences over candidates, and used these estimated preferences to match employers with well-suited graduating seniors from the University of Pennsylvania. This allowed them to estimate employers true preferences over candidates---revealed through their ratings of the hypothetical resumes---without spamming employers with fake resumes.

Neither of these papers discuss the actual ``algorithm'' used, nor do they tell subjects what it is.\footnote{For example, \citet{kessler2019resume} tell employers that they use ``a newly developed machine-learning algorithm to identify candidates who would be a particularly good fit for your job based on your evaluations.''} Thus, as in our paper, it is potentially manipulable or possible that subjects thought that it was not incentive compatible. In contrast to these papers, our design allows us to test for untruthful reporting through a variety of measures summarized above, and we find no evidence of it.

\section{Discussion} \label{sec:discussion}
Using a method that allows us to elicit incompleteness directly, we find that about 40\% of subjects have incomplete preferences over simple money lotteries. This manifests most prominently as incomparabilities between lotteries that are relatively ``far apart'' from one another. Incompleteness contributes to standard behavioral anomalies such as intransitivities and preference reversals. Our results suggest incompleteness stems mainly from imprecise tastes rather than imprecise beliefs. 

How should one interpret the incompleteness that subjects reveal directly, or the potential underlying incompleteness that they do not reveal? There are a few alternate interpretations of our data that one could take. First, there are some theories of choice that assume intransitivities must reflect incompleteness; since we do not find that all intransitivities can be explained through incompleteness, under these theories, there must be residual incompleteness that we cannot detect \citep{mandler2005incomplete, nishimura2016incomplete}. This suggests a higher prevalence of incompleteness in the population---up to 94\% of our subjects. However, these preference reversals can also result from stochastic choice, which we cannot differentiate in our data.


Furthermore, it is possible that subjects have incomplete preferences in this space but develop heuristics or procedures that enable them to complete their preferences. We cannot detect this so we cannot rule out this interpretation, and indeed we believe this to be a reasonable hypothesis. Under this interpretation, the reported incompleteness we observe should be interpreted as situations where subjects are unable to find a way to complete their preferences given the heuristics and procedures they have developed. The relationship between heuristics and incompleteness is an interesting open question. 

In addition, and related to our discussions above, the incompleteness we measure in this design is in some sense a \textit{sophisticated} incompleteness, where subjects must know that they don't know their preference. For example, if we assume intransitivities and preference reversals reflect incompleteness as discussed above, we would conclude that subjects are only aware of a fraction of their underlying incompleteness. Other elicitation mechanisms in the literature potentially require less sophistication at the expense of clean interpretation as incompleteness (e.g., randomization). The extent to which individuals are sophisticated about their incompleteness is another interesting open question. 

Finally, the question remains as to how one should interpret the levels of incompleteness we see, especially as it pertains to our unusual elicitation methodology. One can think about this question from two different perspectives: at the level of a given individual or at the level of a given comparison. At the individual-level, we find a very stable proportion of individuals---about 40\%---report incompleteness in our subjective and objective treatments. In our deliberately-complex treatment, this increases to 76\%. This suggests a lower bound of at least three-quarters of subjects trust our elicitation mechanism and are willing to report incompleteness, but many of these subjects do not have incomplete preferences in our standard simple binary choices.

At a comparison-level, it is difficult to say how the level of incompleteness would change in different environments. As noted, we find more incompleteness in complex questions. This suggests that the ``magnitude'' of incompleteness in individuals' preferences is underestimated by the simple questions we use. For example, one could imagine that choices involving compound lotteries, lotteries with more outcomes, or multi-dimensional objects would reveal more incompleteness. From this perspective, it is perhaps surprising that incompleteness remains when we strip away all uncertainty about probabilities and outcomes.

Our results leave open a number of interesting questions. As we discuss in reviewing related papers in the literature, there are other methods of identifying incompleteness, and different measures can lead to different conclusions on preferences. It would be interesting to understand the compare these methodologies to understand the extent to which they measure the same uncertainty in preferences. 

Finally, it would be interesting to understand better how individuals complete their incomplete preference. In our Forced treatment, individuals are asked to make a choice even if their preferences are incomplete. Understanding this completion process better could help interpret standard choice data and potentially could allow for identification of incompleteness even when individuals are unable to report incompleteness directly. Along these lines, it would be very interesting to identify any neurological or biological indicators of completeness.

\newpage
\bibliography{ref}

\begin{thebibliography}{33}
\newcommand{\enquote}[1]{``#1''}
\expandafter\ifx\csname natexlab\endcsname\relax\def\natexlab#1{#1}\fi

\bibitem[\protect\citeauthoryear{Agranov, Healy, and Nielsen}{Agranov
  et~al.}{2021}]{agranov2021stable}
\textsc{Agranov, M., P.~J. Healy, and K.~Nielsen} (2021): \enquote{Stable
  Randomization,} \emph{Working Paper}.

\bibitem[\protect\citeauthoryear{Agranov and Ortoleva}{Agranov and
  Ortoleva}{2017}]{agranov2017randomization}
\textsc{Agranov, M. and P.~Ortoleva} (2017): \enquote{Stochastic choice and
  preferences for randomization,} \emph{Journal of Political Economy}, 40--68.

\bibitem[\protect\citeauthoryear{Agranov and Ortoleva}{Agranov and
  Ortoleva}{2020}]{agranov2020ranges}
---\hspace{-.1pt}---\hspace{-.1pt}--- (2020): \enquote{Ranges of
  Randomization,} \emph{Working Paper}.

\bibitem[\protect\citeauthoryear{Aumann}{Aumann}{1962}]{aumann1962completeness}
\textsc{Aumann, R.} (1962): \enquote{Utility Theory Without the Completeness
  Axiom,} \emph{Econometrica}, 30, 445--462.

\bibitem[\protect\citeauthoryear{Baillon, Halevy, and Li}{Baillon
  et~al.}{Forthcoming}]{halevy2022randomize}
\textsc{Baillon, A., Y.~Halevy, and C.~Li} (Forthcoming): \enquote{Randomize at
  your own Risk: on the Observability of Ambiguity Aversion,}
  \emph{Econometrica}.

\bibitem[\protect\citeauthoryear{Bayrak and Hey}{Bayrak and
  Hey}{2017}]{bayrak2017reversals}
\textsc{Bayrak, O.~K. and J.~D. Hey} (2017): \enquote{Expected Utility Theory
  with Imprecise Probability Perception: Explaining Preference Reversals,}
  \emph{Applied Economics Letters}, 906--910.

\bibitem[\protect\citeauthoryear{Bayrak and Hey}{Bayrak and
  Hey}{2020}]{bayrak2020imprecision}
---\hspace{-.1pt}---\hspace{-.1pt}--- (2020): \enquote{Understanding Preference
  Imprecision,} \emph{Journal of Economic Surveys}, 154--174.

\bibitem[\protect\citeauthoryear{Becker, DeGroot, and Marschak}{Becker
  et~al.}{1964}]{becker1964bdm}
\textsc{Becker, G.~M., M.~H. DeGroot, and J.~Marschak} (1964):
  \enquote{Measuring Utility by a Single-Response Sequential Method,}
  \emph{Behavioral Science}, 9, 226--236.

\bibitem[\protect\citeauthoryear{Bewley}{Bewley}{2002}]{bewley2002knightian}
\textsc{Bewley, T.~F.} (2002): \enquote{Knightian Decision Theory. Part I,}
  \emph{Decisions in Economics and Finance}, 25, 79--110.

\bibitem[\protect\citeauthoryear{Cettolin and Riedl}{Cettolin and
  Riedl}{2019}]{cettolin2019incomplete}
\textsc{Cettolin, E. and A.~Riedl} (2019): \enquote{Revealed Preferences Under
  Uncertainty: Incomplete Preferences and Preferences for Randomization,}
  \emph{Journal of Economic Theory}, 181, 547--585.

\bibitem[\protect\citeauthoryear{Cohen, Jaffray, and Said}{Cohen
  et~al.}{1985}]{cohen-jaffray-said_85}
\textsc{Cohen, M., J.-Y. Jaffray, and T.~Said} (1985): \enquote{Individual
  Behavior under Risk and under Uncertainty: An Experimental Study,}
  \emph{Theory and Decision}, 18, 203--228.

\bibitem[\protect\citeauthoryear{Cohen, Jaffray, and Said}{Cohen
  et~al.}{1987}]{cohen-jaffray-said_87}
---\hspace{-.1pt}---\hspace{-.1pt}--- (1987): \enquote{Experimental Comparison
  of Individual Behavior under Risk and under Uncertainty for Gains and for
  Losses,} \emph{Organizational Behavior and Human Decision Processes}, 39,
  1--22.

\bibitem[\protect\citeauthoryear{Costa-Gomes, Cueva, Gerasimou, and
  Teji\v{s}\v{c}\'{a}k}{Costa-Gomes et~al.}{2021}]{costagomes2021deferral}
\textsc{Costa-Gomes, M., C.~Cueva, G.~Gerasimou, and M.~Teji\v{s}\v{c}\'{a}k}
  (2021): \enquote{Choice, Deferral and Consistency,} \emph{Quantitative
  Economics}, forthcoming.

\bibitem[\protect\citeauthoryear{Cubitt, Navarro-Martinez, and Starmer}{Cubitt
  et~al.}{2015}]{cubitt2015imprecision}
\textsc{Cubitt, R.~P., D.~Navarro-Martinez, and C.~Starmer} (2015): \enquote{On
  Preference Imprecision,} \emph{Journal of Risk and Uncertainty}, 50, 1--34.

\bibitem[\protect\citeauthoryear{Danz, Vesterlund, and Wilson}{Danz
  et~al.}{2022}]{danz2022belief}
\textsc{Danz, D., L.~Vesterlund, and A.~J. Wilson} (2022): \enquote{Belief
  Elicitation and Behavioral Incentive Compatibility,} \emph{American Economic
  Review, Forthcoming}, 112, 28511--83.

\bibitem[\protect\citeauthoryear{Dubra, Maccheroni, and Ok}{Dubra
  et~al.}{2004}]{dubra2004euincomplete}
\textsc{Dubra, J., F.~Maccheroni, and E.~A. Ok} (2004): \enquote{Expected
  Utility Theory without the Completeness Axiom,} \emph{Journal of Economic
  Theory}, 115, 118--33.

\bibitem[\protect\citeauthoryear{Eliaz and Ok}{Eliaz and
  Ok}{2006}]{eliaz2006indifference}
\textsc{Eliaz, K. and E.~A. Ok} (2006): \enquote{Indifference or
  Indecisiveness? Choice-Theoretic Foundations of Incomplete Preferences,}
  \emph{Games and Economic Behavior}, 56, 61--86.

\bibitem[\protect\citeauthoryear{Fox and Tversky}{Fox and
  Tversky}{1995}]{fox1995comparative}
\textsc{Fox, C.~R. and A.~Tversky} (1995): \enquote{Ambiguity Aversion and
  Comparative Ignorance,} \emph{The Quarterly Journal of Economics}, 110,
  585--603.

\bibitem[\protect\citeauthoryear{Galaabaatar and Karni}{Galaabaatar and
  Karni}{2013}]{galaabaatar2013seu}
\textsc{Galaabaatar, T. and E.~Karni} (2013): \enquote{Subjective Expected
  Utility with Incomplete Preferences,} \emph{Econometrica}, 81, 255--284.

\bibitem[\protect\citeauthoryear{Gilboa, Maccheroni, Marinacci, and
  Schmeidler}{Gilboa et~al.}{2010}]{gilboa2010objective}
\textsc{Gilboa, I., F.~Maccheroni, M.~Marinacci, and D.~Schmeidler} (2010):
  \enquote{Objective and subjective rationality in a multiple prior model,}
  \emph{Econometrica}, 78, 755--770.

\bibitem[\protect\citeauthoryear{Giustinelli, Manski, and Molinari}{Giustinelli
  et~al.}{2019}]{giustinelli2019precise}
\textsc{Giustinelli, P., C.~F. Manski, and F.~Molinari} (2019):
  \enquote{Precise of Imprecise Probabilities? Evidence from Survey Response on
  Late-onset Dementia,} \emph{NBER Working Paper}.

\bibitem[\protect\citeauthoryear{Karni}{Karni}{2018}]{karni2018mechanism}
\textsc{Karni, E.} (2018): \enquote{A Mechanism for Eliciting Second-Order
  Beliefs and the Inclination to Choose,} \emph{American Economic Journal:
  Microeconomics}.

\bibitem[\protect\citeauthoryear{Karni}{Karni}{2021}]{karni2022randomincomplete}
---\hspace{-.1pt}---\hspace{-.1pt}--- (2021): \enquote{Incomplete Preferences
  and Random Choice Behavior: Axiomatic Characterizations,} \emph{Working
  Paper}.

\bibitem[\protect\citeauthoryear{Karni and Vier\o}{Karni and
  Vier\o}{2020}]{karni2020comparative}
\textsc{Karni, E. and M.-L. Vier\o} (2020): \enquote{Comparative
  Incompleteness: Measurement, Behavioral Manifestations and Elicitation,}
  \emph{Working Paper}.

\bibitem[\protect\citeauthoryear{Kessler, Low, and Sullivan}{Kessler
  et~al.}{2019}]{kessler2019resume}
\textsc{Kessler, J.~B., C.~Low, and C.~D. Sullivan} (2019):
  \enquote{Incentivized Resume Rating: Eliciting Employer Preferences without
  Deception,} \emph{American Economic Review}, 109, 3713--3144.

\bibitem[\protect\citeauthoryear{Krajbich, Camerer, and Rangel}{Krajbich
  et~al.}{2017}]{krajbich2017neurometrically}
\textsc{Krajbich, I., C.~Camerer, and A.~Rangel} (2017): \enquote{Exploring the
  Scope of Neurometrically Informed Mechanism Design,} \emph{Games and Economic
  Behavior}, 101, 49--62.

\bibitem[\protect\citeauthoryear{Mandler}{Mandler}{2005}]{mandler2005incomplete}
\textsc{Mandler, M.} (2005): \enquote{Incomplete Preferences and Rational
  Intransitivity of Choice,} \emph{Games and Economic Behavior}, 255--277.

\bibitem[\protect\citeauthoryear{Mas-Colell, Whinston, and Green}{Mas-Colell
  et~al.}{1995}]{mwg}
\textsc{Mas-Colell, A., M.~D. Whinston, and J.~R. Green} (1995):
  \emph{Microeconomic Theory}, New York: Oxford University Press.

\bibitem[\protect\citeauthoryear{Nishimura and Ok}{Nishimura and
  Ok}{2016}]{nishimura2016incomplete}
\textsc{Nishimura, H. and E.~A. Ok} (2016): \enquote{Utility Representation of
  an Incomplete and Nontransitive Preference Relation,} \emph{Journal of
  Economic Theory}, 166, 164--185.

\bibitem[\protect\citeauthoryear{Ok}{Ok}{2002}]{ok2002uincomplete}
\textsc{Ok, E.~A.} (2002): \enquote{Expected Utility Theory without the
  Completeness Axiom,} \emph{Journal of Economic Theory}, 104, 429--449.

\bibitem[\protect\citeauthoryear{Ok, Ortoleva, and Riella}{Ok
  et~al.}{2012}]{ok2012incomplete}
\textsc{Ok, E.~A., P.~Ortoleva, and G.~Riella} (2012): \enquote{Incomplete
  Preferences Under Uncertainty: Indecisiveness in Beliefs versus Tastes,}
  \emph{Econometrica}, 80, 1791--1808.

\bibitem[\protect\citeauthoryear{Rigotti and Shannon}{Rigotti and
  Shannon}{2005}]{rigottishannon2005}
\textsc{Rigotti, L. and C.~Shannon} (2005): \enquote{Uncertainty and Risk in
  Financial Markets,} \emph{Econometrica}, 73, 203--243.

\bibitem[\protect\citeauthoryear{Wang, Filiba, and Camerer}{Wang
  et~al.}{2010}]{Wang2010DOSE}
\textsc{Wang, S., M.~Filiba, and C.~Camerer} (2010): \enquote{Dynamically
  Optimized Sequential Experimentation (DOSE) for Estimating Economic
  Preference Parameters,} \emph{mimeo}.

\end{thebibliography}

\clearpage
\appendix 

\section{Additional Tables and Figures}\label{sec:appendixResults}

\begin{table}[!htb]
    \centering
    \begin{tabular}{lc}
    \hline \hline 
    \multicolumn{2}{c}{Ever Report Incompleteness} \\ \hline 
        Female & 0.359** \\
            & (0.155) \\
        Age & -0.00471 \\
        & (0.00665) \\
        Education & 0.0897\\
        & (0.595) \\
        SES & -0.0374\\
        & (0.0466) \\
        Investment & -0.0261 \\
        & (0.165) \\ 
        Constant & -0.353\\
        & (0.313) \\
        \hline \hline 
    \end{tabular}
    \caption{Demographic Predictors of Incompleteness}
    \label{reg:demographics}
    \footnotesize{\underline{Note:} This is a probit regression where the dependent variable is an indicator for ever reporting incompleteness. Standard errors are in parentheses.}
\end{table}    

\begin{table}[!htb]
    \centering
    \begin{tabular}{cc}
    \hline \hline 
    Payment if Not Verb (\$)     & Payment if Verb (\$) \\ \hline 
5 &	19 \\
5 &	16 \\
6 &	18 \\
7 &	10 \\
7 &	16 \\
8 &	17 \\
7 &	19 \\
8 &	13 \\
6 &	12 \\
9 &	14 \\
9 &	9 \\
9 &	11 \\
10 &	8 \\
10 &	13 \\
10 &	6 \\
13 &	3 \\
16 &	3 \\
11 &	5 \\
12 &	1 \\
11 &	10 \\
12 &	6 \\
14 &	2 \\
14 &	4 \\
17 &	1 \\
12 &	8 \\
 \hline \hline 
\hline \hline 
    \end{tabular}
    \caption{List of All Lotteries }
    \label{tab:listoflotteries}
    \footnotesize{\underline{Note:} (9, 11) and (14, 2) were the two reference lotteries. }
\end{table}

\begin{figure}[!htb]
    \centering
    \includegraphics[width=\textwidth]{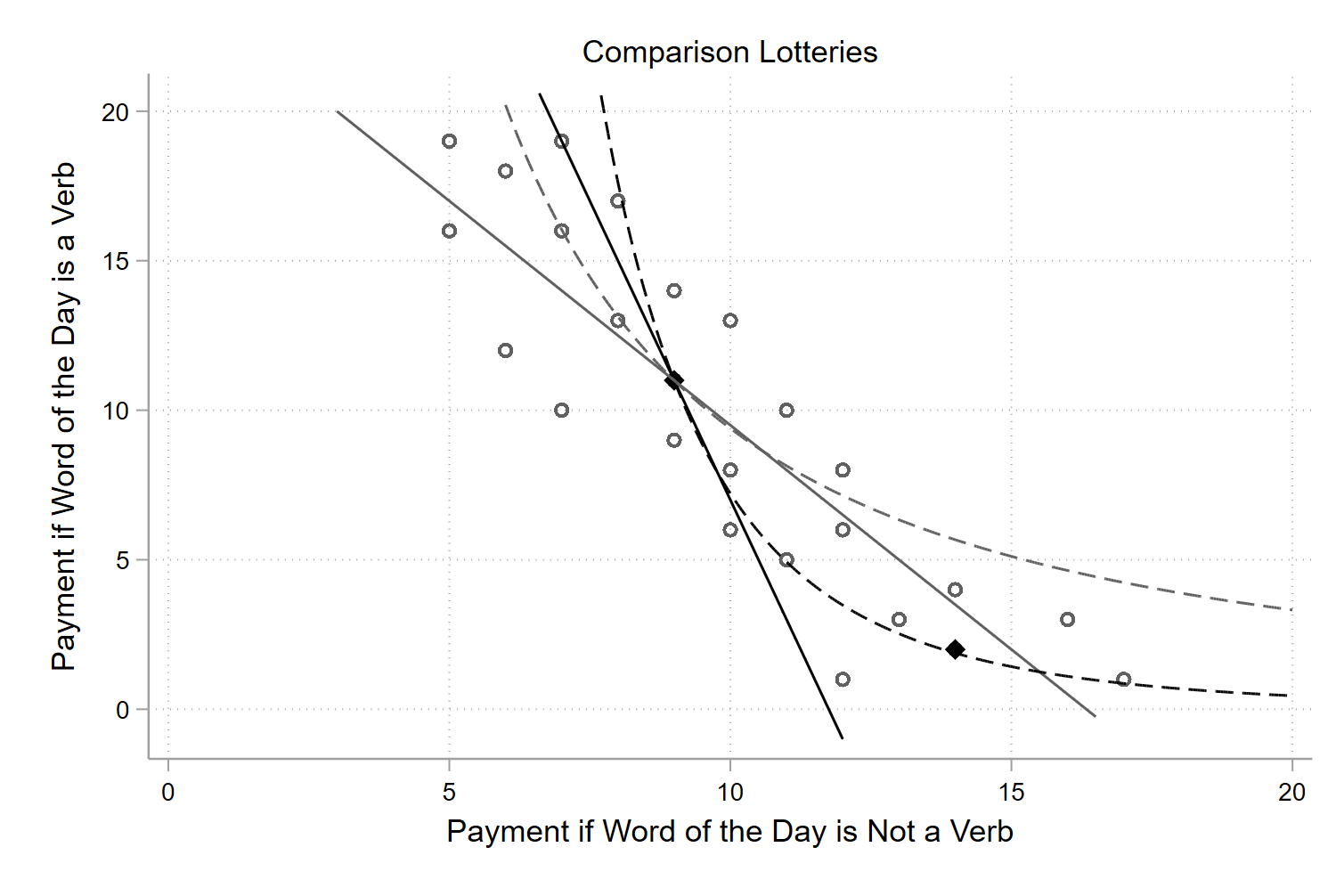}
    \caption{Comparison Lotteries Calibrated Against Utility Functions}
    \label{fig:911functions}
    \footnotesize{\underline{Note:} Solid lines show linear indifference curves while dashed lines show log-utility. The black lines show $pr(verb)=0.2$ while grey lines show $pr(verb)=0.4$.}
\end{figure}

\begin{figure}[!htb]
    \centering
    \includegraphics[width=\textwidth]{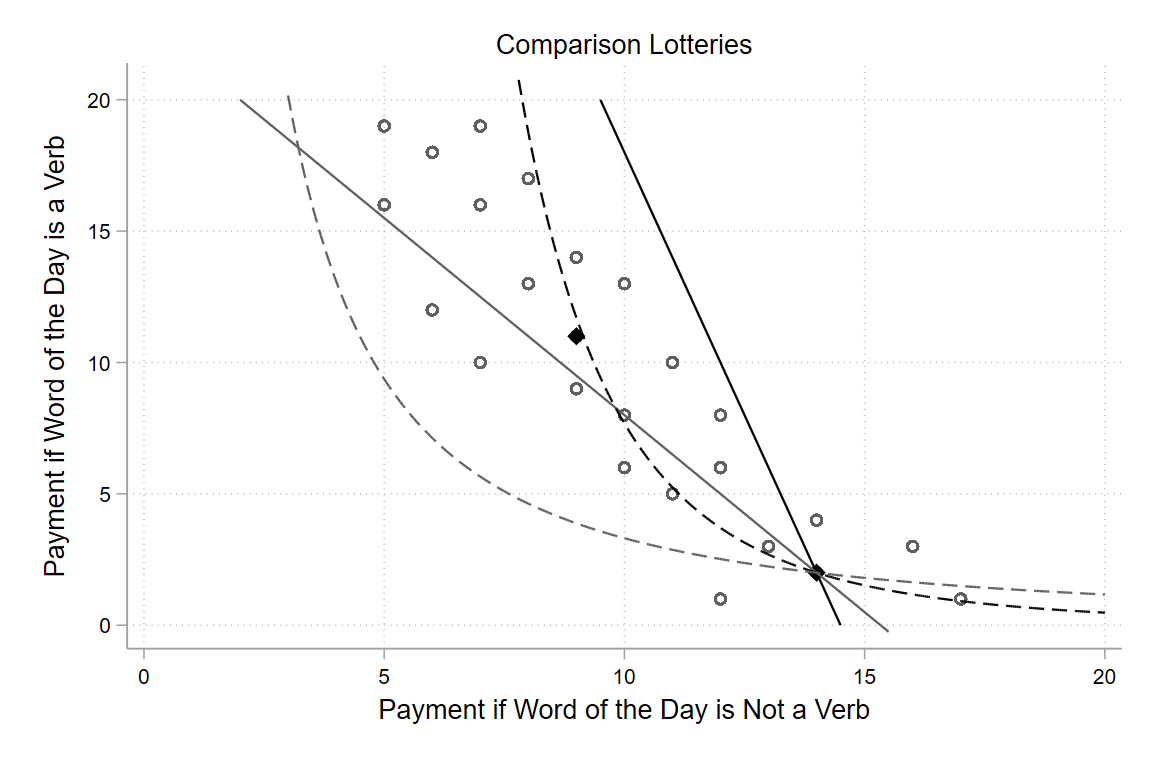}
    \caption{Comparison Lotteries Calibrated Against Utility Functions}
    \label{fig:142functions}
    \footnotesize{\underline{Note:} Solid lines show linear indifference curves while dashed lines show log-utility. The black lines show $pr(verb)=0.2$ while grey lines show $pr(verb)=0.4$.}
\end{figure}

\begin{figure}
    \centering
    \includegraphics[width=\textwidth]{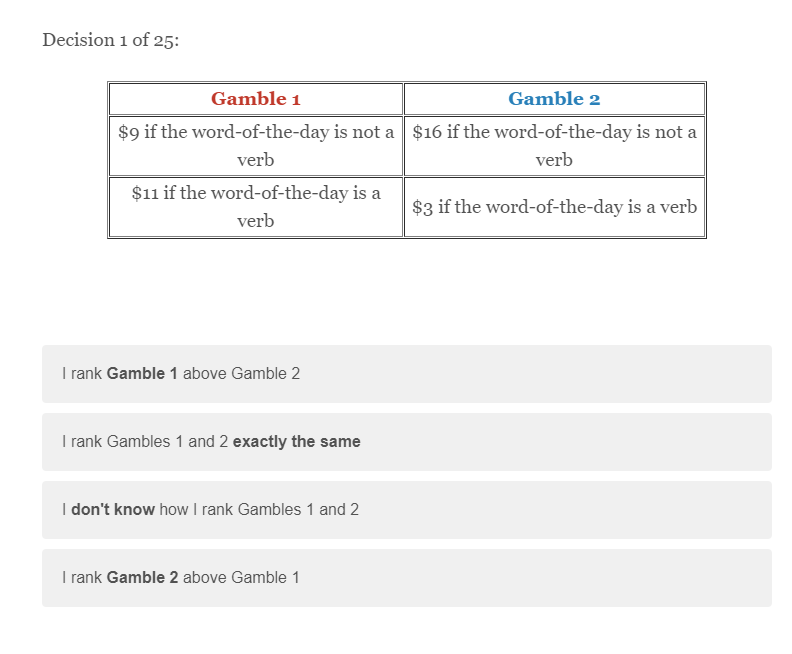}
    \caption{Screenshot of Decision Screen in Non-Forced Treatment}
    \label{fig:screenshot}
    \footnotesize{\underline{Note:} Decisions in the Forced treatment looked exactly the same, without the indifferent and incomplete answer options.}
\end{figure}

\clearpage 
\section{Details on Algorithms}\label{sec:appendix_algorithms}
Our main data contains responses from 639 participants recruited using two different elicitation algorithms. These two algorithms potentially offer different reporting incentives and have different theoretical properties, but we show in Table~\ref{tab:algorithmcomparison} that subjects' responses were the same regardless of the exact incentive mechanism.

We designed our first algorithm, that we call the ``Set Construction Algorithm,'' to map out the theoretical objects of better-than and worse-than sets. We start with a ``better-than set'' and a ``worse-than set'' that each contain 10 randomly-selected lotteries from the space of all possible lotteries $(x,y)$, with $x, y \in[0,20]$. When a subject reports that $(x, y)$ is strictly preferred to a reference lottery, we replace one of the lotteries in the better-than set with $(x+i, y+i), i \in [1,5]$.\footnote{In most of our data, $i=1$. However, technically there are two comparison lottery pairs for which $(x_1 + 1, y_1+1) = (x_2, y_2)$, so one of the lotteries that could be put into the better-than or worse-than set was itself a comparison lottery. Because of this, in later treatments, we randomly select $i$ and put a restriction to prevent this. We do not believe this makes a difference in practice. } When a subject reports that the reference lottery is strictly preferred to $(x,y)$, we replace one of the lotteries in the worse-than set with $(x-i, y-i)$. When a subject reports indifference between a reference lottery and $(x,y)$, we replace one of the lotteries in the better-than set \textit{and} one in the worse-than set as described above. When a subject reports incompleteness, we do not change the better than and worse than sets.

If a subject were paid from this procedure, then we randomly selected to pay them from the better-than set or worse-than set with equal chance. If we randomly selected the worse-than set, then the subject would receive the reference lottery as payment (since they preferred the reference lottery to any lottery in the worse-than set).\footnote{Technically, this means that subjects do not have a strict preference to report when a lottery is worse than the reference lottery, since they will be paid the reference lottery anyway. We could have introduced strict incentive by replacing a lottery from the worse-than set and \textit{also} changing the better-than set, for example, by removing any lotteries in the better-than set that were worse than the current lottery. We avoided this for the additional complication, and so that there was no risk of the better-than set becoming empty. Instead, we included lotteries that were dominated by the reference lottery so that we could identify whether subjects report when the reference lottery is strictly better. As we discuss in Section~\ref{sec:manipulation}, 90\% of subjects do.} If we randomly selected the better-than set, then we would randomly select one of the lotteries in the better-than set, and the subject would receive this lottery as payment.

In theory, this procedure is desirable for a few reasons. First, given that subjects are never paid for any of the exact choices $p$ vs. $q$ that they make throughout the experiment, we are never forced to complete their preference between $p$ and $q$ comparisons in which they report incompleteness. Second, the procedure described above allows for estimated preferences to be incomplete. We are simply constructing better-than and worse-than sets, and the union of these sets need not (and in fact does not) span the entire space of lotteries. Third, in this simple set construction procedure, we make no functional form or other parametric assumptions, so choices only inform preferences through dominance. Finally, the question we ultimately pay subjects is one where we are ``sure,'' under dominance, that they have complete preferences.

The one major drawback of this type of procedure is that the questions subjects answer influence the possible lottery choices they could be paid, exactly because of this last feature where we only pay subjects for a choice where they have complete preferences. In particular, the lotteries that are replaced into the better-than and worse-than sets are a function (via dominance) of the questions that a subject answers with strict preference or indifference. As a result, this algorithm is not incentive compatible for all possible beliefs and preferences. For example, if a subject thought that the better than set contained ``very good'' lotteries to start with, then they would not want to indicate that any comparison lotteries were preferred to the reference lottery, since this would replace a ``very good'' lottery with an inferior one. There is no reason for subjects to hold this belief and it is not accurate---the better-than and worse-than sets start out with randomly-generated lotteries, and we state this in the additional information about the algorithm---but we cannot entirely rule out such beliefs.\footnote{We needed the sets to start with randomly-selected lotteries to give subjects an incentive to answer questions where they had a strict preference. Otherwise, a subject could, for example, answer one question and be guaranteed that lottery for payment.}

Given these pros and cons, we re-ran our experiment using a different---and potentially more standard---incentive mechanism, that we call the ``MLE Algorithm.'' Here, we fixed a single payment question for all participants, $(14, 2)$ vs. $(9, 5)$, which was not one of the questions that subjects faced in the experiment. Using a standard maximum likelihood approach, we took all of the questions in which a subject reported strict preference or indifference to estimate a CRRA utility function for each subject who received a bonus payment. Incomparabilities simply did not enter into this estimation. We then calculated whether the subject would prefer $(14, 2)$ or $(9, 5)$ given their estimated utility function, and we paid them based on this prediction.

Here, the incentives are, loosely, that reporting strict preferences or indifference can help the estimation procedure form a more precise estimate of the utility function. Since subjects do not know what question is being paid, if their response could change the estimated utility function, then there are questions where this change in estimate could lead to different payment lotteries. When subjects are not sure of their preference in a given comparison, reporting incompleteness prevents this comparison from potentially biasing the estimation.

The drawbacks of this type of procedure are, in some sense, the opposite of the Set Construction Algorithm. Here, we fix a payment question ex-ante and we force a complete preference in this comparison. It could be the case that subjects actually have incomplete preferences in this question, but this procedure does not allow for that. Furthermore, we also have to use a specific functional form for the estimation. In theory, it could be that subjects have complete but non-CRRA preferences, which could lead to biases in reporting.

\begin{table}[!htb]
    \centering
    \begin{tabular}{c|cccc}
    \hline \hline 
    Reference Lottery & Prefer Reference & Prefer Comparison & Indifferent & Incomplete \\ \hline 
     Set Construction &  34.2\% & 55.0\% & 8.2\% & 2.6\% \\
     MLE  &  33.4\% & 56.2\% & 7.7\% & 2.7\% \\ 
     \hline \hline 
    \end{tabular}
    \caption{Aggregate Choice Data}
    \footnotesize{\underline{Note:} Subjects made 25 comparisons for each reference lottery. The table presents the percentage of subjects who preferred the reference lottery, preferred the comparison lottery, were indifferent between the two, and were unable to compare the two, aggregated across subjects.}
    \label{tab:algorithmcomparison}
\end{table}

Table~\ref{tab:algorithmcomparison} shows that subjects' responses are statistically indistinguishable across these algorithms (Fisher's exact $p=0.249$). The same percentage of subjects clicked to learn more about the algorithm ($p=0.829$), and the choices among those who read the algorithm details are also statistically indistinguishable $(p=0.098))$. Given that these two algorithms are very different from one another, and that the theoretical pros and cons are quite opposite, we find this compelling evidence that the details of the underlying algorithm do not affect subjects' responses.

In running our MLE Algorithm treatment, we also recruited 166 additional subjects and incentivized them under the same algorithm but exogenously provided them with the information about the algorithm. That is, rather than choosing to learn the information or not by clicking a button, we showed this information to all participants. Specifically, we tell subjects the following:
\begin{quote}
    We will use maximum-likelihood estimation to estimate a constant relative risk aversion utility function. Maximum likelihood estimation will find the constant relative risk aversion parameter such that your choices are most probable under this model. Then, we will use that model to predict what you would choose in another question, and will pay you based on this prediction. The choices that you make will help this estimation procedure to choose the parameter that best fits your preferences. If you say that you "do not know" which gamble you prefer in a question, we will not use this question in our maximum likelihood estimation. We will use only the questions where you know which gamble you prefer.
\end{quote}
We link Wikipedia pages for maximum likelihood estimation, constant relative risk aversion, and utility functions.

We find a slight reduction in the percentage of subjects who report incompleteness in this treatment (31\% vs. 41\%, $p=0.071$). We find it implausible that this reflects subjects reading and optimally responding to the MLE and CRRA detailed incentives. Instead, this seems in line with \citet{danz2022belief} who find that providing detailed information on incentives can lead to subjects distorting their responses. In support of this, we find that, among subjects who ever report incompleteness, they report the same percentage of incomplete comparisons at a question level ($p=0.414$) regardless of whether information was exogenously provided. Thus, it appears that some subjects are ``scared off'' or confused by the elicitation details, so too much information can cause a reduction in incompleteness on the extensive margin but not on the intensive margin. 

Given that we find the details of the algorithm do not affect behavior, we believe that this type of elicitation falls under the umbrella of mechanisms that are \textit{behaviorally} incentive compatible \citep{danz2022belief} even if not theoretically incentive compatible. This seems natural given the complexity of understanding the incentive to manipulate. We liken this to other methodologies such as dynamically optimized sequential experimentation (DOSE, \citealp{Wang2010DOSE}) and the other papers in the literature that have used preference estimation without entirely fixing subjects' beliefs about the incentives \citep{krajbich2017neurometrically, kessler2019resume}.

\end{document}